\documentclass[10pt,conference]{IEEEtran}
\usepackage{cite}
\usepackage{amsmath,amssymb,amsfonts}
\usepackage{algorithmic}
\usepackage{graphicx}
\usepackage{textcomp}
\usepackage[hyphens]{url}
\usepackage{fancyhdr}
\usepackage[table,dvipsnames]{xcolor}
\usepackage{multirow}
\usepackage{enumitem}
\usepackage{lipsum}
\usepackage{longtable}
\usepackage{makecell}
\usepackage{float}
\usepackage{subcaption}
\usepackage{soul}
\usepackage[linesnumbered,ruled,vlined]{algorithm2e}
\usepackage{amsmath}
\usepackage{amssymb}
\usepackage[T1]{fontenc}
\usepackage{hyperref}

\newcommand{\hpcayear}{2025}

\newcommand{\hpcasubmissionnumber}{60}
\title{MLPerf Power: Benchmarking the Energy Efficiency of Machine Learning Systems from \textmu Watts to MWatts for Sustainable AI}

\def\hpcacameraready{}

\newcommand\hpcaauthors{
\begin{minipage}[t]{0.9\textwidth} 
\centering 
Arya Tschand$^{1*}$ Arun Tejusve Raghunath Rajan$^{2*}$ 
Sachin Idgunji$^{3*}$ Anirban Ghosh$^{3}$ Jeremy Holleman$^{4}$ 
Csaba Kiraly$^{5}$ Pawan Ambalkar$^{6}$ Ritika Borkar$^{3}$ 
Ramesh Chukka$^{7}$ Trevor Cockrell$^{6}$ Oliver Curtis$^{8}$ 
Grigori Fursin$^{9}$ Miro Hodak$^{10}$ Hiwot Kassa$^{2}$ 
Anton Lokhmotov$^{11}$ Dejan Miskovic$^{3}$ Yuechao Pan$^{12}$ 
Manu Prasad Manmathan$^{7}$ Liz Raymond$^{6}$ Tom St. John$^{13}$ 
Arjun Suresh$^{14}$ Rowan Taubitz$^{8}$ Sean Zhan$^{8}$ 
Scott Wasson$^{15}$ David Kanter$^{15}$ Vijay Janapa Reddi$^{1}$
\end{minipage}
\vspace{8px}}

\newcommand\hpcaaffiliation{
\begin{minipage}[t]{0.9\textwidth} 
\centering 
$^{*}$Equal contribution\hspace{1px}
$^{1}$Harvard University\hspace{1px}
$^{2}$Meta\hspace{1px}
$^{3}$NVIDIA\hspace{1px}
$^{4}$UNC Charlotte / Syntiant\hspace{1px}
$^{5}$Codex\hspace{1px}
$^{6}$Dell\hspace{1px}
$^{7}$Intel\hspace{1px}
$^{8}$SMC\hspace{1px}
$^{9}$FlexAI / cTuning\hspace{1px}
$^{10}$AMD\hspace{1px}
$^{11}$KRAI\hspace{1px}
$^{12}$Google\hspace{1px}
$^{13}$Decompute\hspace{1px}
$^{14}$GATE Overflow\hspace{1px}
$^{15}$MLCommons
\end{minipage}}

\author{
  \ifdefined\hpcacameraready
    \IEEEauthorblockN{\hpcaauthors{}}
      \IEEEauthorblockA{
        \hpcaaffiliation{} \\
      }
  \else
    \IEEEauthorblockN{\normalsize{HPCA \hpcayear{} Submission
      \textbf{\#\hpcasubmissionnumber{}}} \\
      \IEEEauthorblockA{
        Confidential Draft \\
        Do NOT Distribute!!
      }
    }
  \fi 
}

\fancypagestyle{camerareadyfirstpage}{%
  \fancyhead{}
  
  \fancyhead[C]{
    \ifdefined\aeopen
    \parbox[][12mm][t]{13.5cm}{\hpcayear{} IEEE International Symposium on High-Performance Computer Architecture (HPCA)}    
    \else
      \ifdefined\aereviewed
      \parbox[][12mm][t]{13.5cm}{\hpcayear{} IEEE International Symposium on High-Performance Computer Architecture (HPCA)}
      \else
      \ifdefined\aereproduced
      \parbox[][12mm][t]{13.5cm}{\hpcayear{} IEEE International Symposium on High-Performance Computer Architecture (HPCA)}
      \else
      \parbox[][0mm][t]{13.5cm}{\hpcayear{} IEEE International Symposium on High-Performance Computer Architecture (HPCA)}
    \fi 
    \fi 
    \fi 
    \ifdefined\aeopen 
      \includegraphics[width=12mm,height=12mm]{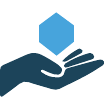}
    \fi 
    \ifdefined\aereviewed
      \includegraphics[width=12mm,height=12mm]{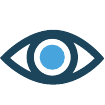}
    \fi 
    \ifdefined\aereproduced
      \includegraphics[width=12mm,height=12mm]{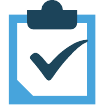}
    \fi
  }
  \fancyfoot[C]{}
}
\begin{document}
\maketitle

\ifdefined\hpcacameraready 
  \thispagestyle{camerareadyfirstpage}
  \pagestyle{empty}
\else
  \thispagestyle{plain}
  \pagestyle{plain}
\fi

\newcommand{\hpcaheight}{0mm}
\ifdefined\eaopen
\renewcommand{\hpcaheight}{12mm}
\fi

\begin{abstract}

Rapid adoption of machine learning (ML) technologies has led to a surge in power consumption across diverse systems, from tiny IoT devices to massive datacenter clusters. Benchmarking the energy efficiency of these systems is crucial for optimization, but presents novel challenges due to the variety of hardware platforms, workload characteristics, and system-level interactions. This paper introduces MLPerf® Power, a comprehensive benchmarking methodology with capabilities to evaluate the energy efficiency of ML systems at power levels ranging from microwatts to megawatts. Developed by a consortium of industry professionals from more than 20 organizations, coupled with insights from academia, MLPerf Power establishes rules and best practices to ensure comparability across diverse architectures. We use representative workloads from the MLPerf benchmark suite to collect 1,841 reproducible measurements from 60 systems across the entire range of ML deployment scales. Our analysis reveals trade-offs between performance, complexity, and energy efficiency across this wide range of systems, providing actionable insights for designing optimized ML solutions from the smallest edge devices to the largest cloud infrastructures. This work emphasizes the importance of energy efficiency as a key metric in the evaluation and comparison of the ML system, laying the foundation for future research in this critical area. We discuss the implications for developing sustainable AI solutions and standardizing energy efficiency benchmarking for ML systems.

\end{abstract}
\section{Introduction}
\label{sec:intro}

\begin{figure}
    \centering
    \includegraphics[width=\columnwidth]{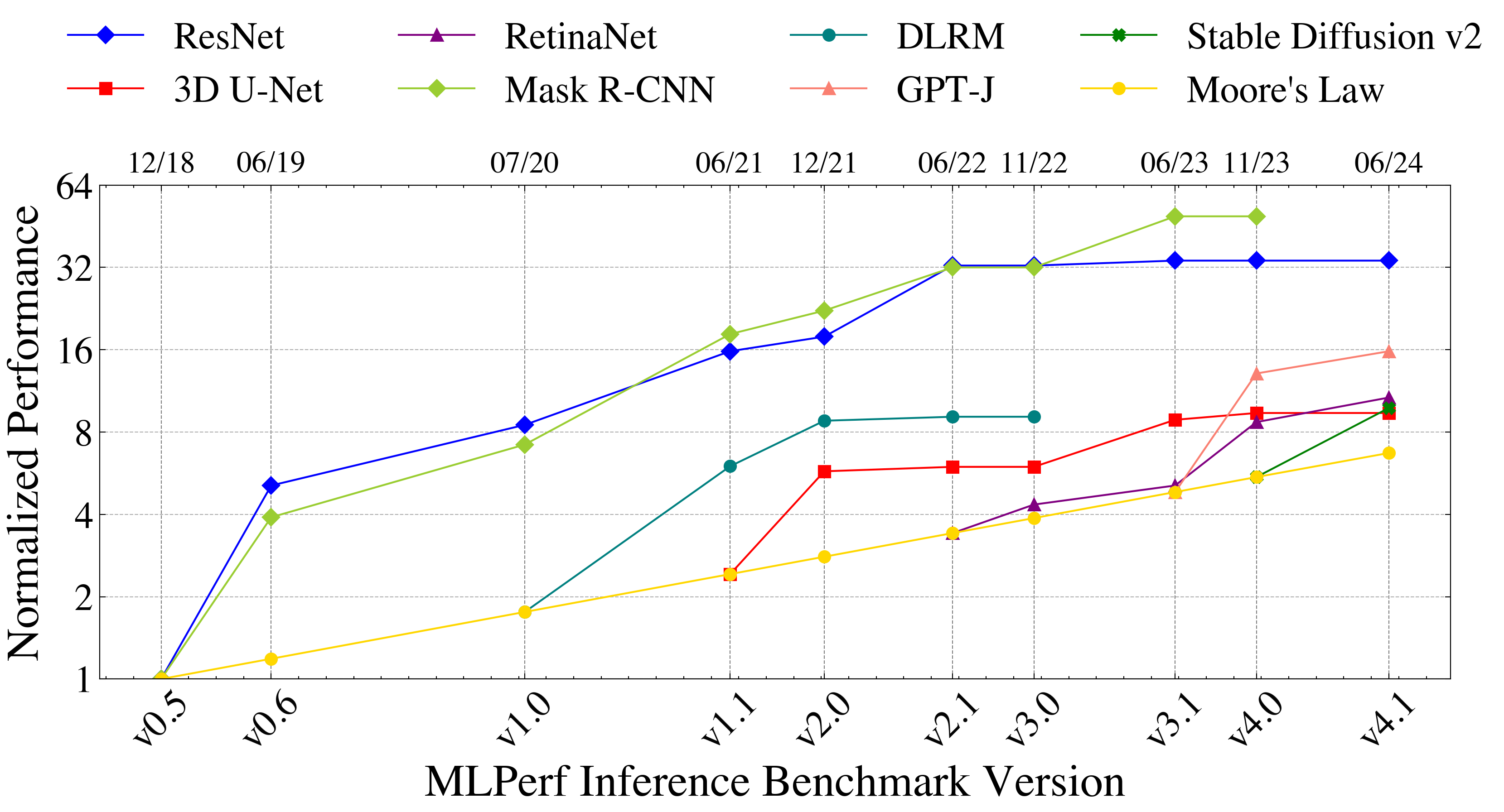}
    \caption{MLPerf performance improvements have outpaced Moore's Law. This trend highlights the rapid evolution of AI systems, prompting the development of MLPerf Power to address emerging concerns over their energy efficiency.}
    \label{fig:mlperf_performance}
\end{figure}

In recent years, machine learning (ML) technologies have transformed a wide range of fields, from high-performance computing centers to edge devices and small IoT systems. The exceptional improvement in ML performance is clearly demonstrated by the MLPerf Training benchmark results~\cite{mlperftraining}, as shown in Figure~\ref{fig:mlperf_performance}. Since the debut of this benchmark in 2018, performance has soared, showing an incredible 32-fold increase. This rapid pace of advancement is not limited to established benchmarks; newly introduced benchmarks also show consistent performance enhancements. Such exponential growth has led us to surpass Moore's law, which highlights the extraordinary pace of innovation in ML systems driven by hardware, algorithm, and software enhancements.

The advancement in ML capabilities has coincided with a notable increase in concerns about the power consumption of ML systems~\cite{sevilla2022compute, desislavov2021compute, samsi2023words, stojkovic2024towards, prakash2023tinyml}. As the ecological impact and operational expenses of these systems continue to increase, they have become a significant concern for both industry and academia~\cite{wu2022sustainable, rillig2023risks, vahdatnew, towardssustainable}. The critical nature of this challenge highlights the need for a standardized and thorough benchmarking approach to measure the energy efficiency of ML systems, from small IoT devices to large-scale cloud infrastructures. This method must also guarantee accurate reproducibility, facilitate meaningful comparisons, and encourage advances in energy-efficient ML technologies.

However, measuring and comparing the energy efficiency of ML systems presents unique challenges that differentiate them from traditional computing systems. Unlike conventional general-purpose workloads, ML systems span various hardware platforms, from specialized accelerators to general-purpose processors~\cite{lim2021f1}, each with their own power consumption characteristics~\cite{lee2021overview, madmax}. Moreover, ML workloads exhibit diverse computational patterns and resource requirements, such as high data parallelism, memory intensity, and communication-bound operations, which differ significantly from traditional CPU-centric applications. The complexity is further compounded by system-level interactions, such as data movement between memory hierarchies and communication overheads in distributed systems, which have a more profound impact on ML systems' energy efficiency than traditional setups. These characteristics emphasize the need for a comprehensive benchmarking methodology that can accurately capture and analyze energy efficiency across different hardware configurations, workload types, and system scales.

In addition, the wide spectrum of ML systems, ranging from IoT devices to expansive data centers, presents a formidable challenge in developing a universally applicable power measurement methodology. This diversity requires a scalable approach capable of accurately capturing power consumption characteristics across heterogeneous domains. Although the fundamental principles of power measurement remain consistent, the specific techniques and considerations vary significantly depending on the scale and nature of the ML system under evaluation, which we elaborate in this paper. 

These diverse requirements emphasize the need for a comprehensive and adaptable power measurement methodology. To address the multifaceted challenges of measuring and comparing the energy efficiency of ML systems, we developed MLPerf Power, a comprehensive benchmarking methodology. This initiative is the result of collaborative efforts among 20 organizations actively participating in the MLPerf Training~\cite{mlperftraining} and Inference~\cite{mlperfinference} benchmarks, which have long served as the industry standard for measuring ML system performance. MLPerf Power extends this framework, recognizing that performance metrics alone are insufficient to enable the development of sustainable AI systems for the future.

MLPerf Power represents a concerted industry-wide effort to establish a standardized approach for evaluating the power consumption of ML systems across an unprecedented range of energy levels, from microwatts in resource-constrained IoT devices to megawatts in high-performance computing clusters. Using the collective expertise and industry experience of its contributors, MLPerf Power defines a consistent set of benchmarking rules, measurement techniques, and reporting guidelines. This framework enables accurate and fair comparisons between various ML systems, irrespective of their hardware configurations or workload characteristics.

We provide a detailed overview of the MLPerf Power methodology and its application across various ML systems. The methodology is designed to account for the unique aspects of ML workloads, including power consumption implications of data preparation, model training, and inference phases. MLPerf Power places a strong emphasis on transparency and reproducibility, mandating detailed disclosure of system specifications and power measurement setups. We describe the MLPerf Power methodology's foundational design principles and key components, exploring the benchmarking rules that establish a level playing field for diverse ML systems, the measurement techniques ensuring accurate power consumption data across different scales, and the reporting guidelines promoting transparency and enabling meaningful comparisons. 

The application of the MLPerf Power methodology to a diverse range of ML systems has yielded powerful insights into energy efficiency trends and optimization strategies. Our comprehensive analysis reveals both promising advances and emerging challenges in the search for more sustainable AI solutions. Energy efficiency trends across MLPerf versions show a general positive trajectory, but exhibit recent plateaus in improvements for certain workloads and scales. In particular, older and better optimized workloads such as \texttt{ResNet} and \texttt{RNN-T} have reached a point of diminishing returns in recent versions. In contrast, newer workloads like \texttt{GPT-J} and \texttt{Llama2} demonstrate rapid improvements, with energy efficiency improving over 100$\times$ in about a year between versions, driven by intense commercial interest and high-impact optimizations.

Our study of datacenter-scale training submissions for models such as Llama2-70b highlights the complex trade-offs between system scale, energy efficiency, and performance. As the number of accelerators scales by a factor of 64$\times$, we observe that energy consumption increases only 3.8$\times$, while training time plummets by 93\%. This nonlinear relationship underscores the importance of considering both energy and time when evaluating large-scale ML systems. Furthermore, to provide actionable insights, we dissect these high-level energy efficiency improvements, isolating the contributions of various optimization strategies. Through a case study on progressive versions of neural network ASICs with comparable software stacks, we demonstrate that while performance improvements plateau at around 1.5$\times$, significant reductions in power consumption enable a remarkable 4$\times$ improvement in overall energy efficiency. In contrast, our analysis of software isolated improvements in comparable hardware submissions reveals that negligible performance sacrifices (less than 1\%) can yield substantial energy efficiency gains of up to 28\%. Furthermore, we illustrate how more aggressive quantization techniques, when carefully applied to maintain benchmark accuracy targets, can increase energy efficiency by up to 70\%.

As ML continues to evolve and expand, the importance of energy efficiency in ML systems cannot be overstated. As newer and more complex generative models are consuming orders of magnitude more energy per inference, our findings emphasize the importance of considering energy efficiency as a first-class metric in system design and optimization. The insights and best practices presented in this work will help guide the design, deployment, and optimization of energy-efficient ML systems across various domains, from edge devices to cloud-scale infrastructures.

In summary, this paper makes the following contributions:

\begin{itemize}
\item \textbf{Standardization of power measurement for ML systems across the industry:} MLPerf Power introduces a comprehensive, industry-wide benchmarking methodology for measuring power consumption in ML systems. This enables fair comparisons across diverse hardware platforms and workloads, from tiny IoT devices to large datacenter clusters. MLPerf Power measures the true power consumption during standardized workloads to track improvements in ML hardware and software optimizations, and promote transparency across the industry.

\item \textbf{First large-scale study with practical guidance on energy efficiency in ML systems:} We present a comprehensive analysis of over 1,800 energy measurement results across diverse ML systems in production. We also evaluate the real-world impact of various optimization techniques, including hardware improvements, software optimizations, and quantization. This dual approach offers insights into the current state of energy efficiency in ML and provides actionable guidance for practitioners and researchers to make informed decisions about optimization strategies for their specific use cases.



\item \textbf{Industry-wide analysis of ML system energy efficiency trends:} The paper presents a longitudinal study of energy efficiency improvements across multiple generations of MLPerf Power submissions from various industry players. This analysis provides insights into the state of energy efficiency in commercial ML systems, highlighting areas of progress and identifying opportunities for further optimization. It offers a unique perspective on how different sectors of the industry (datacenter, edge, tiny) are addressing energy efficiency challenges.

\end{itemize}


\section{Background, Motivation and Prior Work}

\begin{table*}[t!]
\setlength{\textfloatsep}{5pt plus 1.0pt minus 2.0pt}
\caption{Related work. We identify several key requirements that are necessary for an industry-standard ML power benchmark.}
\label{table:sample}
\centering
\begin{tabular}{|c|c|c|c|c|c|c|c|c|c|c|c|c|}
\hline
\multirow{2}{*}{Category} & \multirow{2}{*}{Paper} & \multicolumn{4}{c|}{\textcolor{red}{Power Measurement}} & \multicolumn{3}{c|}{\textcolor{blue}{Diverse ML Benchmarking}} & \multicolumn{4}{c|}{\textcolor{orange}{Real-World Evaluation, Trends, Impact}} \\
\cline{3-13}
& & \textcolor{red}{Req. 1} & \textcolor{red}{Req. 2} & \textcolor{red}{Req. 3} & \textcolor{red}{Req. 4} & \textcolor{blue}{Req. 5} & \textcolor{blue}{Req. 6} & \textcolor{blue}{Req. 7} & \textcolor{orange}{Req. 8} & \textcolor{orange}{Req. 9} & \textcolor{orange}{Req. 10} & \textcolor{orange}{Req. 11} \\
\hline
Benchmark & SPEC Power~\cite{lange2009identifying} & \textcolor{green}{\checkmark} & \textcolor{red}{X} & \textcolor{red}{X} & \textcolor{green}{\checkmark} & \textcolor{gray}{-} & \textcolor{gray}{-} & \textcolor{gray}{-} & \textcolor{gray}{-} & \textcolor{gray}{-} & \textcolor{gray}{-} & \textcolor{gray}{-} \\
\hline
Benchmark & Green500~\cite{feng2007green500} & \textcolor{green}{\checkmark} & \textcolor{red}{X} & \textcolor{red}{X} & \textcolor{green}{\checkmark} & \textcolor{gray}{-} & \textcolor{gray}{-} & \textcolor{gray}{-} & \textcolor{gray}{-} & \textcolor{gray}{-} & \textcolor{gray}{-} & \textcolor{gray}{-} \\
\hline
Benchmark & Top500~\cite{dongarra1997top500} & \textcolor{green}{\checkmark} & \textcolor{red}{X} & \textcolor{red}{X} & \textcolor{green}{\checkmark} & \textcolor{gray}{-} & \textcolor{gray}{-} & \textcolor{gray}{-} & \textcolor{gray}{-} & \textcolor{gray}{-} & \textcolor{gray}{-} & \textcolor{gray}{-} \\
\hline
Measurement & Carbon~\cite{patterson2021carbon} & \textcolor{red}{X} & \textcolor{red}{X} & \textcolor{green}{\checkmark} & \textcolor{red}{X} & \textcolor{green}{\checkmark} & \textcolor{green}{\checkmark} & \textcolor{green}{\checkmark} & \textcolor{green}{\checkmark} & \textcolor{red}{X} & \textcolor{red}{X} & \textcolor{green}{\checkmark} \\
\hline
Measurement & Systematic~\cite{henderson2020towards} & \textcolor{green}{\checkmark} & \textcolor{green}{\checkmark} & \textcolor{red}{X} & \textcolor{red}{X} & \textcolor{red}{X} & \textcolor{red}{X} & \textcolor{green}{\checkmark} & \textcolor{green}{\checkmark} & \textcolor{red}{X} & \textcolor{red}{X} & \textcolor{red}{X} \\
\hline
Measurement & Carbontracker~\cite{anthony2020carbontracker} & \textcolor{red}{X} & \textcolor{red}{X} & \textcolor{green}{\checkmark} & \textcolor{red}{X} & \textcolor{green}{\checkmark} & \textcolor{red}{X} & \textcolor{red}{X} & \textcolor{green}{\checkmark} & \textcolor{red}{X} & \textcolor{red}{X} & \textcolor{red}{X} \\
\hline
Measurement & AI Training~\cite{wang2020aitraining} & \textcolor{green}{\checkmark} & \textcolor{green}{\checkmark} & \textcolor{red}{X} & \textcolor{red}{X} & \textcolor{red}{X} & \textcolor{red}{X} & \textcolor{green}{\checkmark} & \textcolor{red}{X} & \textcolor{red}{X} & \textcolor{red}{X} & \textcolor{red}{X} \\
\hline
Measurement & USB~\cite{libutti2020usb} & \textcolor{red}{X} & \textcolor{red}{X} & \textcolor{green}{\checkmark} & \textcolor{green}{\checkmark} & \textcolor{green}{\checkmark} & \textcolor{green}{\checkmark} & \textcolor{green}{\checkmark} & \textcolor{red}{X} & \textcolor{red}{X} & \textcolor{red}{X} & \textcolor{red}{X} \\
\hline
Optimization & Zeus~\cite{you2023zeus} & \textcolor{green}{\checkmark} & \textcolor{red}{X} & \textcolor{red}{X} & \textcolor{red}{X} & \textcolor{red}{X} & \textcolor{red}{X} & \textcolor{green}{\checkmark} & \textcolor{green}{\checkmark} & \textcolor{red}{X} & \textcolor{red}{X} & \textcolor{red}{X} \\
\hline
Trends & Trends DL~\cite{desislavov2021trendsdl} & \textcolor{gray}{-} & \textcolor{gray}{-} & \textcolor{gray}{-} & \textcolor{gray}{-} & \textcolor{gray}{-} & \textcolor{gray}{-} & \textcolor{gray}{-} & \textcolor{green}{\checkmark} & \textcolor{red}{X} & \textcolor{green}{\checkmark} & \textcolor{red}{X} \\
\hline
Trends & Survey~\cite{reuther201survey} & \textcolor{gray}{-} & \textcolor{gray}{-} & \textcolor{gray}{-} & \textcolor{gray}{-} & \textcolor{gray}{-} & \textcolor{gray}{-} & \textcolor{gray}{-} & \textcolor{red}{X} & \textcolor{red}{X} & \textcolor{green}{\checkmark} & \textcolor{red}{X} \\
\hline\hline
All of the above & \textbf{MLPerf Power} & \textcolor{green}{\checkmark} & \textcolor{green}{\checkmark} & \textcolor{green}{\checkmark} & \textcolor{green}{\checkmark} & \textcolor{green}{\checkmark} & \textcolor{green}{\checkmark} & \textcolor{green}{\checkmark} & \textcolor{green}{\checkmark} & \textcolor{green}{\checkmark} & \textcolor{green}{\checkmark} & \textcolor{green}{\checkmark} \\
\hline
\end{tabular}
\vspace{-5mm}
\end{table*}

This section establishes the requirements for benchmarking and measuring the energy efficiency of ML systems and demonstrates how our approach advances the field. We outline the key requirements and then gives an overview of the limitations of existing methods using Table~\ref{table:sample}. MLPerf Power was specifically designed to meet all the identified requirements.

\noindent
\textbf{Power Measurements}
\begin{enumerate}[label=\it Req \arabic*., leftmargin=*]
    \item Compatible with a wide range of platforms and configurations (and their power consumption characteristics).

    \item Account for system-level interactions and shared resources that impact ML (i.e., system energy efficiency, heterogeneous systems accelerator + host + etc.).
    
    \item Adapt measurements to capture power consumption characteristics across different scales of ML systems.
    
    \item Physical power measurements with an analyzer.
\end{enumerate}

\noindent
\textbf{Diverse ML Benchmarking}
\begin{enumerate}[label=\it Req \arabic*., leftmargin=*]
    \setcounter{enumi}{4}
    \item Consistent benchmarking rules and reporting guidelines (for accurate and fair comparisons).
    \item Representative workloads from the industry-standard MLPerf Benchmarking Suite (tiny to datacenter).
    \item Collect data on power consumption and performance metrics (e.g., throughput, latency) at different throughputs and ML model accuracy/quality targets.
\end{enumerate}

\noindent
\textbf{Evaluation, Trends, and Impact}
\begin{enumerate}[label=\it Req \arabic*., leftmargin=*]
    \setcounter{enumi}{7}
    \item Communicate power consumption analysis and optimizations for sustainable AI (patterns and bottlenecks differ significantly across scales of ML systems).
    \item Be driven and audited by the industry, which fosters fair and representative performance evaluations.
    \item Communicate energy efficiency trends (observed across a wide spectrum of ML platform scales, power levels, workload characteristics, hardware configurations, system-level optimizations).
    \item Case studies showcasing the application of MLPerf Power to real-world scenarios on production-grade systems (demonstrate its effectiveness in identifying energy efficiency bottlenecks and guiding the development of more sustainable AI solutions).
\end{enumerate}

MLPerf Power addresses all necessary requirements, combining the strengths of innovative academic approaches with industry-standard best practices. This sets a new benchmark for measuring energy efficiency in machine learning systems.


\section{ML System Power Measurement Challenges}

The goal of MLPerf is to be a system level performance benchmark for machine learning (ML) systems. Therefore, our goal while developing the power measurement methodologies for MLPerf was to develop a full system power measurement methodology that is applicable across the diverse hardware platforms that MLPerf supports and scalable across different MLPerf categories that target different system scales. This section highlights the challenges encountered while developing the MLPerf power measurement methodology.


\subsection{Scale of ML Systems}

\begin{figure}[t!]
    \centering    
    \includegraphics[width=\columnwidth]{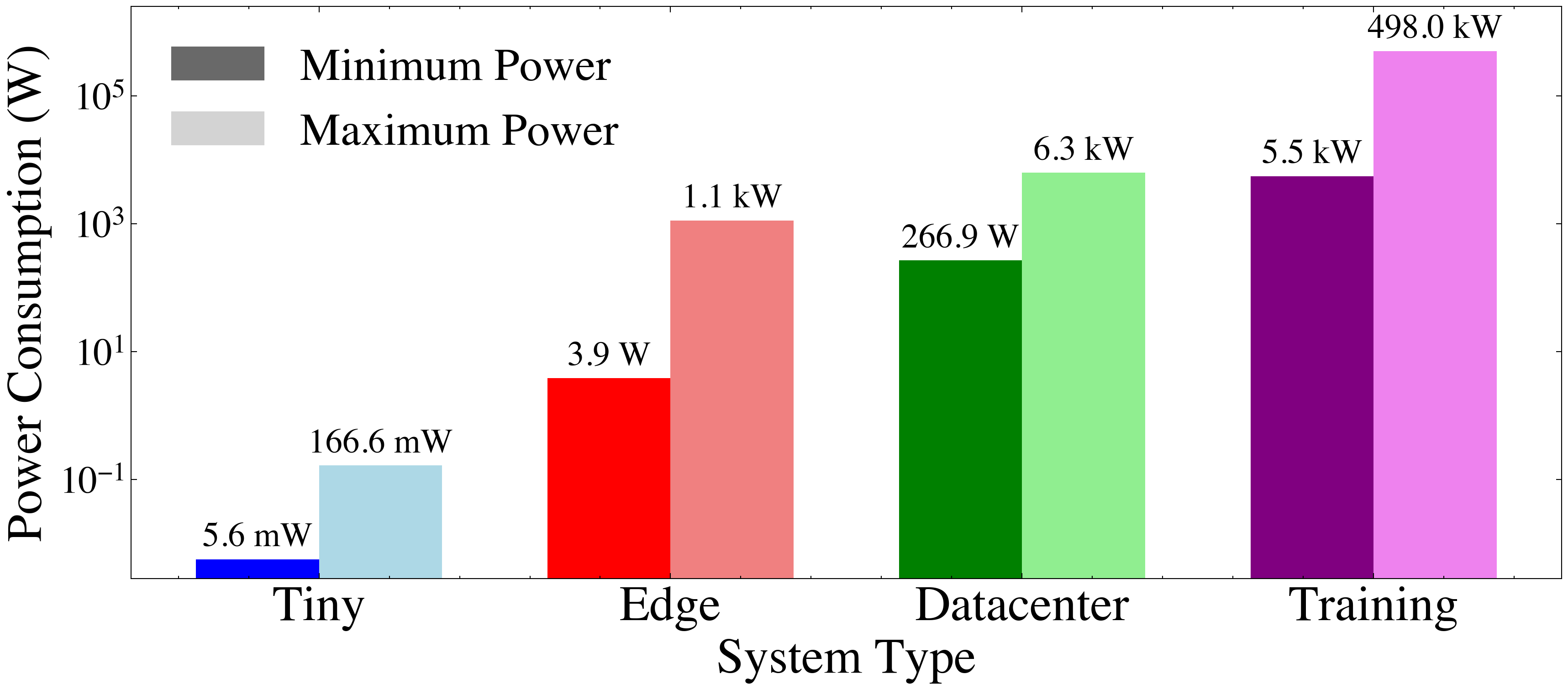}
    \caption{The power consumption range across MLPerf divisions, highlighting the need for scalable power measurement.}
    \label{fig:mlsystems_scale}
\end{figure}

MLPerf Power is designed to stretch from microwatts to megawatts. Figure \ref{fig:mlsystems_scale} samples from the official MLPerf Power submission results to showcase this diverse range of systems, which poses a challenge in creating a unified approach to measuring full system power. MLPerf-Tiny systems operate with a power consumption as low as 5.64 mW. They typically process incoming (sensor) data before entering a low-power standby mode until the next data frame arrives, resulting in duty cycles that frequently fall below 5\%. Consequently, even systems with milliwatt-level peak power consumption often draw just microwatts in real-world average power consumption. At the other end of the scale, explicit power submissions for MLPerf-Training can reach up to 500 kW~\cite{mlperftraining}. However, in large-scale submissions that currently only submit performance measurements, we estimate power consumption around 10 MW in MLPerf-Training and 30 MW in MLPerf-HPC.

To address this diversity in ML applications and benchmarks, it is essential to tailor methodologies for three distinct scales while measuring system power consumption as comprehensively as possible for each scale.

\textbf{Datacenter Training / HPC} To measure the full power of a system, we must account for all components that contribute to training performance. For multi-node training, this includes the compute nodes, interconnect fabric, and any cooling infrastructure. Due to the scale (10K+ GPUs) and nature (secure HPC, on-premise or cloud datacenters) of these systems, it is impractical to use conventional power meters to measure system power. Instead, we use a combination of hardware counters for compute and estimation for switching subsystems.

MLPerf supports a large diversity of submitters, each with their own telemetry and monitoring setup. However, measurement of interconnect power is challenging, as switches may not have as much telemetry to report power and cloud service providers often do not want to reveal their network topology to competitors. To address this, we need to develop a software-based approach that leverages estimation methods and existing telemetry systems, allowing for comprehensive power measurement without compromising security or requiring extensive hardware modifications. 

Measuring cooling power fairly is another challenge. Air-cooled servers have fans whose power is measured when measuring node power, but immersion or direct liquid cooled systems pay their cooling cost at the datacenter level, making it difficult to attribute what fraction was used by the ML job under test. Measuring power consumption from cooling remains future work.




\textbf{Edge} Edge and embedded systems have their own unique challenges when measuring power. For MLPerf-Inference, the Edge devices range from SoCs rated for tens of watts to edge servers that can draw several kilowatts. The inference methodology requires the use of a SPEC-certified power meter such as the Yokogawa WT310. When measuring AC wall power on very low power devices, we noticed that these power meters tend to have a higher error bar due to the high crest factor caused by power adapters under 75W. Measuring full system power for low power devices may also lead to measuring power of sub-systems that cannot be powered off on the system but do not contribute to ML performance – e.g. WiFi transceivers, on-board sensors, idle IO interfaces, etc. Our approach for edge systems involves using external power analyzers, carefully calibrated to capture the full range of power consumption without introducing measurement error.



\textbf{Mobile} MLPerf Mobile benchmarks the performance of mobile ML systems~\cite{janapa2022mlperf}. However, we currently do not support the measurement of power of mobile systems in MLPerf Power due to the complexity and variability of smartphones. Mobile devices comprise a wide range of components, including communication interfaces, sensors, and peripherals, each with significant fluctuations in power consumption patterns and active components\cite{mobilepower2}. Additionally, the battery-powered nature of these devices means they cannot be simply connected to an external power source for measurement without altering normal operation. Thermal throttling further complicates the relationship between power consumption and performance metrics~\cite{pramanik2019power}. We have piloted efforts using DC-based low-power analyzers, but we have not reached the point of standardization. Given these complexities, establishing a fair and transparent benchmark for measuring power consumption in battery-powered devices is a future goal for MLPerf Power.

\textbf{Tiny} Embedded ML systems relying on microcontrollers often function as detectors, recognizing human presence in a room via camera input or identifying spoken wake words (e.g., ``Alexa'' or ``Hey Siri''). Compared to datacenter or edge, they require a different measurement paradigm. The challenge here lies in capturing extremely low power consumption accurately without the measurement setup itself influencing the results. We need specialized micro-power instrumentation and meticulous setup to capture milliwatt-level fluctuations accurately. We also need to implement strategies to prevent parasitic powering, where the system under test might inadvertently draw power from measurement or communication lines.




\subsection{Heterogeneity and Comparability in Power Measurements}

\textbf{Heterogeneous Hardware} Handling diverse hardware platforms presents another layer of complexity. MLPerf submissions span a wide range of architectures, from general-purpose CPUs to specialized AI accelerators, each with unique power consumption characteristics. To ensure fair comparability, we must develop a flexible methodology that can adapt to different hardware configurations while maintaining consistent measurement principles. This includes guidelines for measuring full system power, accounting for both compute and auxiliary components, and standardized reporting formats that capture hardware-specific details relevant to power consumption.

\textbf{Comparability} Maintaining comparability between submissions is crucial to the integrity of the MLPerf Power benchmarks. To address this challenge, we employ a multi-dimensional strategy. First, we have defined stringent criteria for valid power measurements, such as measurement duration, sampling frequencies, and accuracy standards. Second, we have adopted a uniform logging format to ensure power data from various systems can be consistently analyzed. Third, we mandate comprehensive documentation of measurement configurations and any estimation methods employed, facilitating thorough peer evaluation of submissions.

\begin{figure*}[t]
    \centering
    \includegraphics[width=\textwidth]{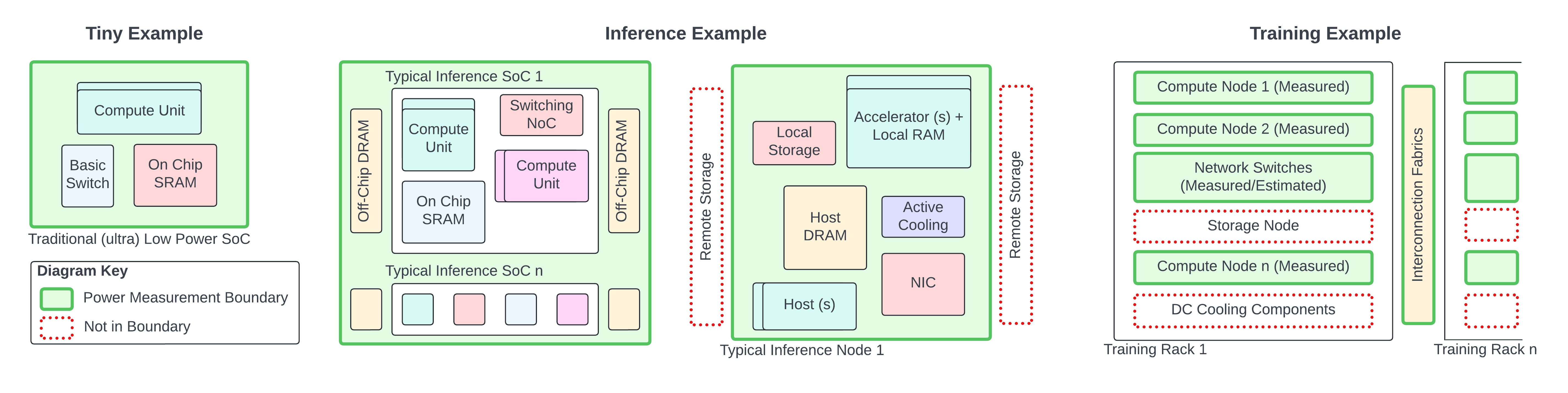}
    \vspace{-5mm}
    \caption{ML system components within the MLPerf Power measurement scope are outlined in green.}
    \label{fig:measurement_scales}
    \vspace{-5mm} 
\end{figure*}

\subsection{Myths and Pitfalls in ML System Power Measurement}
\label{sec:myths}

Before diving into our methodology, we want to address prevalent misconceptions and potential pitfalls in ML system power measurement that contribute to its inherent complexity.

\textbf{Myth \#1: Isolating ML Component Power is Sufficient} A common misconception is that measuring the power consumption of specific ML components, such as accelerators or GPUs, is adequate to assess system efficiency. In reality, overall system power consumption is crucial. Different components are active at various stages of ML workloads with varying duty cycles. For example, during training, accelerators might be active during forward and backward propagation, networking during gradient exchange, and CPUs during data loading. The isolation of component power fails to capture this dynamic interplay, leading to inaccurate representations of energy usage. System-level measurement provides a more comprehensive and realistic view of power consumption.


\textbf{Myth \#2: TDP and PSU Ratings Reflect Power Usage} Another prevalent myth is the reliance on Thermal Design Power (TDP) or Power Supply Unit (PSU) ratings as proxies for power measurement. In reality, TDP only represents the thermal design limit, not the actual power usage in typical workloads. Similarly, PSU ratings include significant margins for power spikes and redundancy, especially in compute servers. These metrics often grossly overestimate actual power consumption, making them poor approximations for real-world usage. Accurate power measurement requires direct monitoring of power consumption during actual ML tasks, rather than relying on these theoretical maximum values.


\textbf{Myth \#3: PUE is Suitable for ML System Efficiency} The use of Power Usage Effectiveness (PUE) to evaluate ML system energy efficiency is a misguided approach. While PUE is crucial for data center efficiency, it is not appropriate for MLPerf methodology for two key reasons. First, PUE reflects datacenter efficiency, not ML system efficiency. Second, verifying PUE claims across diverse submitters is impractical, potentially skewing comparisons between ML systems. Focusing on PUE could mask poor system efficiency with good datacenter efficiency, or vice versa, detracting from MLPerf's goal of tracking ML system improvements. MLPerf's scope is to track improvements in ML system efficiency; building or infrastructure efficiency is beyond its purview.




\textbf{Myth \#4: Precise Power Measurement is Always Possible} There is a common assumption that high-granularity power measurements are universally achievable, which is not always the case. Power measurement granularity is often limited by available infrastructure. Some facilities only measure at the Power Distribution Unit (PDU) level, encompassing multiple nodes, making it challenging to isolate the exact power demands of a specific ML system. This is particularly problematic in cloud environments or shared computing facilities where dynamic job allocation can lead to measurements including idle or unrelated nodes. Additionally, shared cooling resources can make it difficult to isolate the power consumption of specific components. These limitations can result in inaccurate or overly coarse power consumption data, especially in cloud or shared computing environments, highlighting the need for careful consideration of measurement methodologies.


\section{MLPerf Power Measurement Methodology}

The methodology is designed to provide a comprehensive and standardized approach to measuring power consumption in ML systems across a wide range of scales, from microwatts to megawatts. This section outlines the core concepts, metrics, and philosophy behind the MLPerf  power measurement approach.

\subsection{Common Principles Across All Segments}

MLPerf Power is driven by a core set of engineering principles, whether we are dealing with tiny IoT devices or massive datacenters. These principles help ensure that our power measurements are reliable, comparable, and relevant. By adhering to five principles, we are able to provide accurate and meaningful data across a diverse range of systems.

First, a fundamental tenet of MLPerf Power is the emphasis on measuring full system power consumption. This approach recognizes that ML workloads involve various components beyond the primary computing units. In datacenter systems, we account for compute nodes and interconnect fabric. As of today, we do not yet a methodology for including the power for the cooling devices and the storage nodes, but any other drives within the compute blades are accounted for. For edge devices, we consider the entire system-on-chip, including any peripherals that cannot be powered down during operation. In tiny systems, we measure the power of the entire device, including any always-on components. Figure~\ref{fig:measurement_scales} illustrates our measurement for the different scales of systems. Our measurements closely reflect the true energy cost of running ML workloads in real-world scenarios, rather than focusing solely on idealized component-level efficiencies.

Second, ML workloads typically consist of multiple phases: initialization, execution, and result assimilation. The MLPerf Power methodology emphasizes the precision of aligning power measurements with these phases, particularly focusing on the execution phase. This alignment is crucial because it allows for accurate attribution of power consumption to specific parts of the workload, ensures comparability across different systems, and provides insight into the energy efficiency of the core ML computations. To achieve this in tiny systems, we use hardware pins to signal the start and end of inference. In larger systems, we use software-based timestamps to demarcate the execution phase. For training workloads, we parse performance logs to determine the start and stop times of the run.

Third, unlike evaluating overall system performance, the execution phase is more important than end-to-end runtime in power measurement considerations. The execution phase represents the core ML computations, which are typically the most energy-intensive part of the workload. It also provides the most relevant power measurement data to draw actionable insights to improve the energy efficiency of ML systems where it matters most. For inference tasks we measure the power during the actual inference computations. For training tasks, we focus on the power consumed during the training iterations alone. Any of the preceding phases including data assimilation, data setup, offline and online pre-processing steps, are not accounted for. This approach, combined with full system power measurement (as explained in our first principle) and precise alignment with workload phases, forms the foundation of MLPerf Power's methodology across all segments, from the smallest IoT devices to the largest datacenter installations.

Fourth, we ensure comprehensive data collection by implementing a minimum measurement duration. If a workload finishes executing before the 60-second mark, it is run in a loop until this threshold is reached. For workloads that take more than 60 seconds, we collect power data until the workload is completed. We then average the power over the duration of valid samples to report. This approach ensures that we capture the system's full power consumption during execution.

Finally, we employ a unified approach to evaluating energy efficiency on different scales and types of systems. For throughput benchmarks, such as datacenter and offline edge inference~\cite{mlperfinference}, we measure throughput in Samples/s and power consumption in Watts. We calculate energy efficiency as (Samples/s)/(Watts), yielding Samples/Joule. For latency benchmarks such as tiny inference~\cite{mlperftiny}, we use the inverse of energy in 1/Joules. We treat these two metrics as comparable, despite their different units, because the benchmarks fix different execution factors and thus require different performance evaluations. This standardized approach allows for meaningful comparisons across the diverse range of systems.

\subsection{MLPerf Inference Setup and Operation}

Despite the vast difference in the scale of ML inference systems, the core principles of measurement remain consistent, with adaptations made to address the unique challenges of each size of the system. For tiny inference systems, which often function as always-on detectors like wake-word recognizers or presence detectors, the energy per inference is the key metric. These systems align well with the single-stream scenario in MLPerf Inference, which is the time to process one sample. 

\begin{figure*}[t]
    \centering
    \begin{subfigure}[b]{0.30\textwidth}
        \centering
        \includegraphics[width=\textwidth]{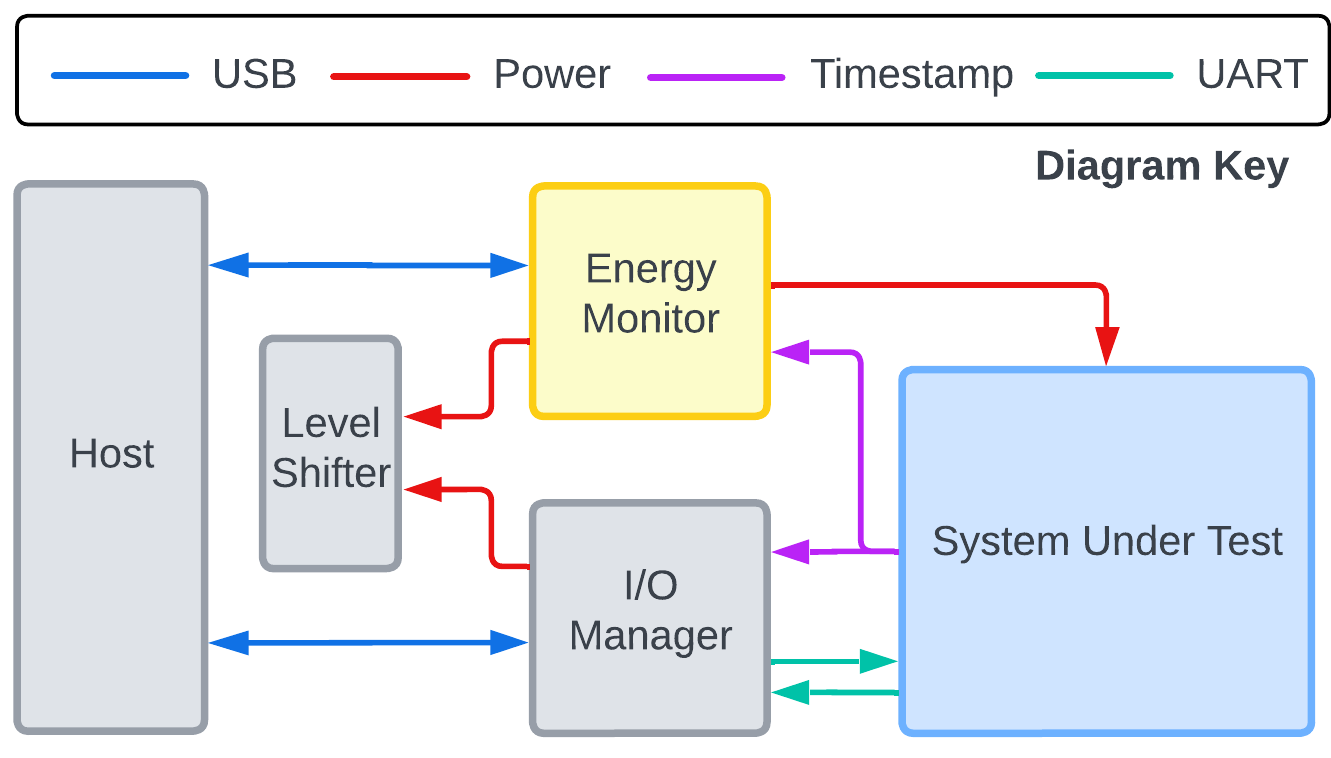}
        \caption{Tiny System}
        \label{fig:tiny_measurement}
    \end{subfigure}
    \begin{subfigure}[b]{0.34\textwidth}
        \centering
        \includegraphics[width=\textwidth]{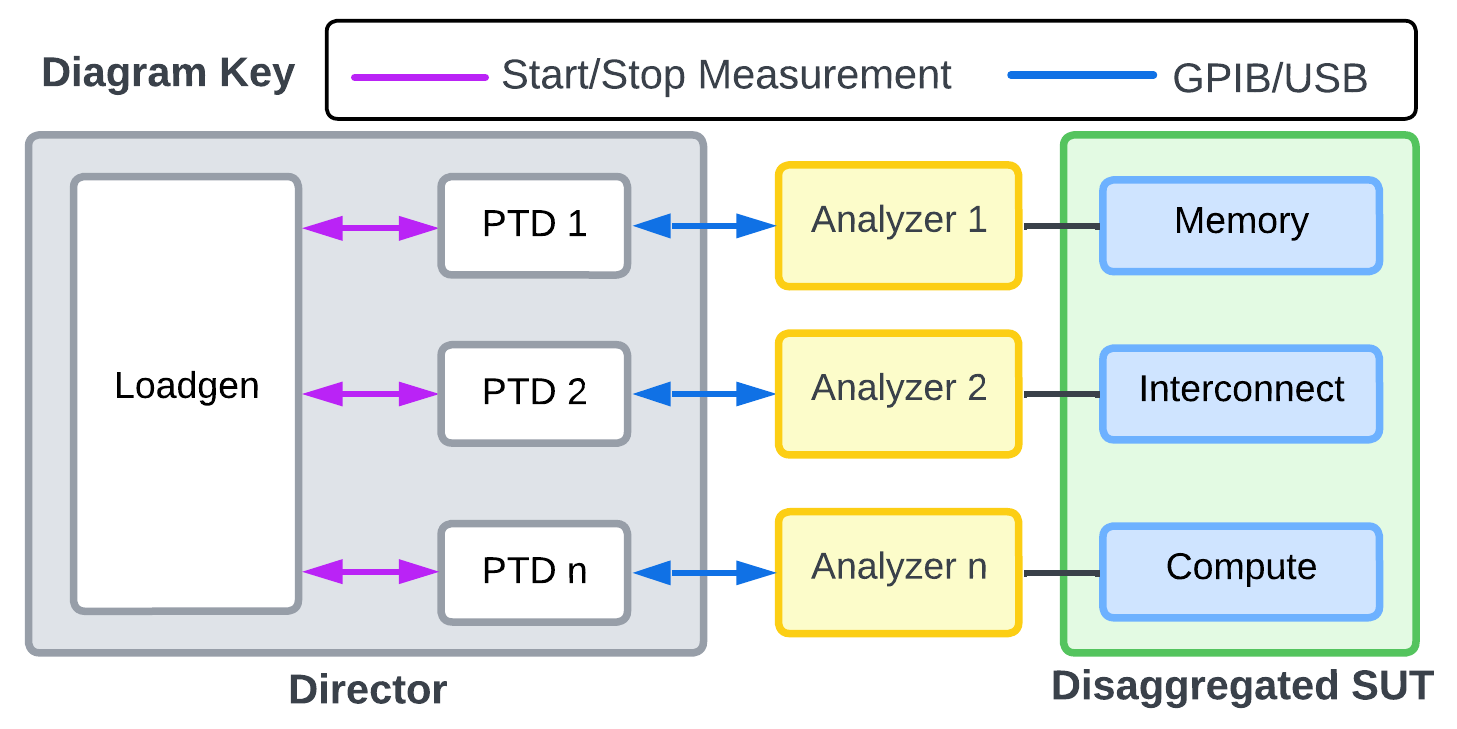}
        \caption{Multi-SUT Inference System}
        \label{fig:multi_sut_measurement}
    \end{subfigure}
    \begin{subfigure}[b]{0.31\textwidth}
        \centering
        \includegraphics[width=\textwidth]{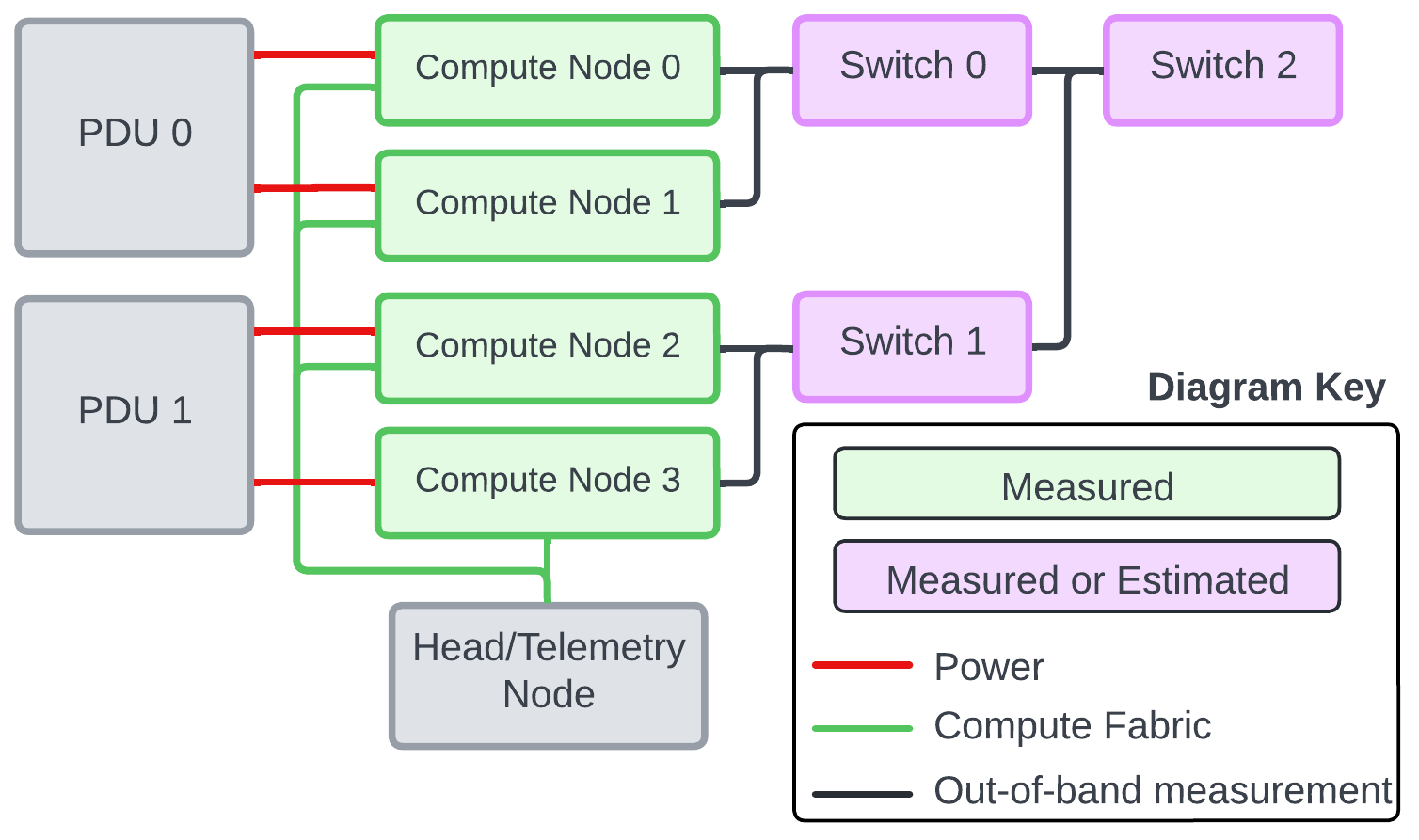}
        \caption{Training System}
        \label{fig:training_measurement}
    \end{subfigure}
    \caption{Measurement diagrams for Tiny, Multi-SUT Inference, and Training systems.}
    \vspace{-5mm}
    \label{fig:scale_measurements}
\end{figure*}

The energy measurement setup for systems in the Tiny division is illustrated in Figure \ref{fig:tiny_measurement}. To accurately measure such low-power devices, we utilize an I/O Manager that sits between the host and the System Under Test (SUT). The I/O Manager is implemented with an Arduino UNO board that captures timing information with resolution that the host (a Windows, Linux, or Mac) cannot. The I/O Manager, along with the level shifter, also electrically isolates the SUT from the host to avoid parasitic powering from the host's signal lines, skewing the measurements. The I/O Manager provides a USB connection to the host and connects to the SUT's Universal Asynchronous Receiver-Transmitter (UART) through an isolating level shifter, ensuring minimal power transfer. 

To measure the energy, the host issues an \textit{infer} command, which is relayed to the SUT via the I/O manager and includes a number of inferences to be performed. The SUT signals the beginning and end of the inference activity by toggling the \textit{Timestamp} pin. This timing information is captured by the external energy monitor along with the current and voltage waveforms and transmitted to the host. The host then integrates the power over the indicated interval and divides by the number of inferences to calculate the energy per inference.

For larger-scale inference systems, we rely on a client-server architecture where the SUT acts as the client, and a separate system, called the Director, serves as the server. This setup allows for more complex measurement scenarios and better control over the benchmarking process. The process begins with an NTP (Network Time Protocol) sync between the SUT and the Director to align their timestamp references. The Director then initiates communication with a SPEC-approved power analyzer through PTD (Power-Thermal Daemon) API calls. Once the connection is established, the Director commands the SUT to start executing the loadgen, which runs the workload of interest. During execution, the Director logs power data from the analyzer, typically in a text file later converted to CSV format for easier processing. This file contains detailed information on the current, voltage, phase, and power usage of the SUT, all with time stamps.

Currently, the SUT collects performance log files that timestamp different phases of execution. To help us improve measurement accuracy, we employ a range mode that involves an initial run to determine the maximum current and voltage levels for a particular workload. Subsequent runs then use fixed analyzer ranges based on these observed peaks, allowing for higher accuracy within that range. For multi-node inference configurations, we adapt our methodology to handle multiple SUTs connected to a single analyzer or multiple analyzers connected to a single SUT, as shown in Figure \ref{fig:multi_sut_measurement}. This flexibility allows us to accurately measure power consumption in more complex distributed inference scenarios.

Across all scales of inference systems, from tiny to datacenter, we focus on measuring the power consumption during the active portion of the workload. By aligning power measurements with the execution phase and employing techniques appropriate to each scale, we ensure accurate and comparable results across the entire spectrum of inference deployments.

\subsection{MLPerf Training and HPC Setup and Operation}

MLPerf Training and HPC~\cite{mlperfhpc} Power Measurement methodology addresses the unique challenges posed by large-scale ML training systems. These systems range from single nodes to clusters comprising thousands of nodes and tens of thousands of accelerators, often located in secure on-premises datacenters, cloud environments, or national laboratories. The scale, complexity, and security requirements of these systems require a different approach compared to inference setups.

At the core of our methodology is a software-based measurement scheme. This approach leverages the existing telemetry infrastructure of large clusters to measure system power, bypassing the impracticality of using separate power analyzers for each node and interconnect component. The scheme is designed to be flexible, accommodating the diverse range of telemetry systems used by different submitters while still adhering to MLPerf's goal of measuring full system power.

Power measurement at the node level is the foundation of our approach, as shown in Figure \ref{fig:training_measurement}. For each participating compute node, power measurements are taken using the submitter's own telemetry systems, such as IPMI~\cite{minyard2006ipmi} or RedFish~\cite{gonccalves2019standard}. We recommend out-of-band measurement techniques to minimize interference with the ML job itself. In cases where individual node measurement is not feasible, power can be measured at the Power Distribution Unit (PDU) level, provided that all nodes supplied by that PDU are involved in the ML job. Submitters are required to thoroughly document the accuracy of their telemetry systems for review.

Interconnect power measurement presents its own set of challenges. For interconnect switches, power can be measured similarly to that of compute nodes. However, in cases where direct measurement is not possible, we allow estimated power values. When estimation is used, submitters must provide detailed documentation of their estimation methodology, ensuring transparency and comparability between submissions.

To ensure consistency and comparability across submissions with different hardware platforms and configurations, we have implemented a standardized data logging process. Submitters are required to convert their node and interconnect logs to a standardized format using the MLPerf Logging Library.

The energy-to-train calculation is a key component of our methodology. Our result summarizer parses both performance and power logs to derive this metric. The process involves aggregating power consumption data for the duration of the ML training process. The summarizer determines the ML training process window by extracting the timestamps of the run start and run stop log lines in the performance log. It then parses each node's log file to extract power samples that fall within this window. The energy for each node is calculated by integrating power samples over the run's time window, and the total energy is computed by summing the energy across all compute and interconnect components. 


\subsection{Validation and Reporting}

The MLPerf Power methodology places a significant emphasis on the validation of measurements and the standardization of reporting. This is done to ensure accuracy, integrity, comparability, and usefulness of the benchmarking results.

\textbf{Quality} Accuracy requirements form the foundation of our validation process. Although we recognize that the diverse range of systems and scales in ML workloads requires different approaches to power measurement, all must meet stringent accuracy standards as described by MLPerf Inference ~\cite{mlperfinference} and Training~\cite{mlperftraining}. For datacenter and high-performance computing systems, we require documentation of the telemetry systems used, including their accuracy specifications and calibration procedures. Edge systems using external power analyzers must employ SPEC-approved devices, known for their high accuracy and reliability. For TinyML, we mandate specialized micro-power instruments with target accuracy ratings~\cite{mlperftiny}.

\textbf{Sampling Rates} We also specify minimum measurement durations and sampling rates to ensure that the power data adequately captures the system's behavior under the benchmark workload. For instance, we require a minimum of 60 seconds of power data collection for most scenarios, with provisions for longer durations in cases of extended workloads. This ensures that temporary fluctuations or anomalies do not unduly influence the reported power consumption.

\textbf{Reporting} To facilitate comparison between diverse submissions, we have developed a standardized reporting format. This format is designed to capture all relevant details of the power measurement setup, system configuration, and benchmark results. Submitters are required to use the MLPerf Logging Library to convert their raw power logs into this standardized format. The reporting template includes fields for system specifications, power measurement methodology, workload details, and actual power and performance metrics.

The HPC and training reporting and submission process is designed to more comprehensive, flexible, and contextually rich. Submitters are required to report power consumption data along with additional information that could affect these measurements, including details on cooling solutions, power management techniques, and environmental conditions during the testing. The process allows for reporting at either the PDU or node level, accommodating different measurement setups. Submitters run the benchmark with their chosen telemetry option, estimate switch power if necessary, and post-process power logs to conform to the MLPerf Logging Library format. For systems where certain components' power consumption must be estimated, such as interconnect power in large-scale training sets, dedicated sections are provided in the reporting format to document the estimation methodology. The submission package must include the processed data in a specified directory structure, along with documentation on telemetry accuracy. This comprehensive approach ensures transparency and comparability between different submissions, while allowing for the flexibility needed to accommodate various system configurations and measurement approaches.

\textbf{Review} Our review process~\cite{mlperfinferencerules, mlperftrainingrules} is rigorous and multilayered, designed to catch any inconsistencies or anomalies in the submissions. Upon receiving a submission, our review committee, composed of experts in ML systems and power measurement, examines the reported data and documentation. They verify that all required information is present, that the reported measurements comply with the accuracy requirements, and that the power consumption figures are consistent with the system specifications and workload characteristics.


Since the ecosystem is still evolving, in cases where novel measurement techniques or system architectures are involved, the review process may include additional steps. This might involve consultations with domain experts or requests for more detailed explanations of the measurement methodology. The goal is to ensure that innovative approaches can be included in the benchmarks while maintaining the high standards of accuracy and comparability that MLPerf is known for.

Transparency is a key principle of our validation and reporting process. All accepted submissions are made publicly available, along with their full documentation. This allows community scrutiny and fosters trust in the benchmark results. It also provides valuable data for researchers and engineers working to improve the energy efficiency of ML systems~\cite{cm4mlperf}.





\section{Results/Evaluation}

\begin{figure*}[t!]
    \centering
    \begin{subfigure}[b]{0.32\textwidth}
        \centering
        \includegraphics[width=\textwidth]{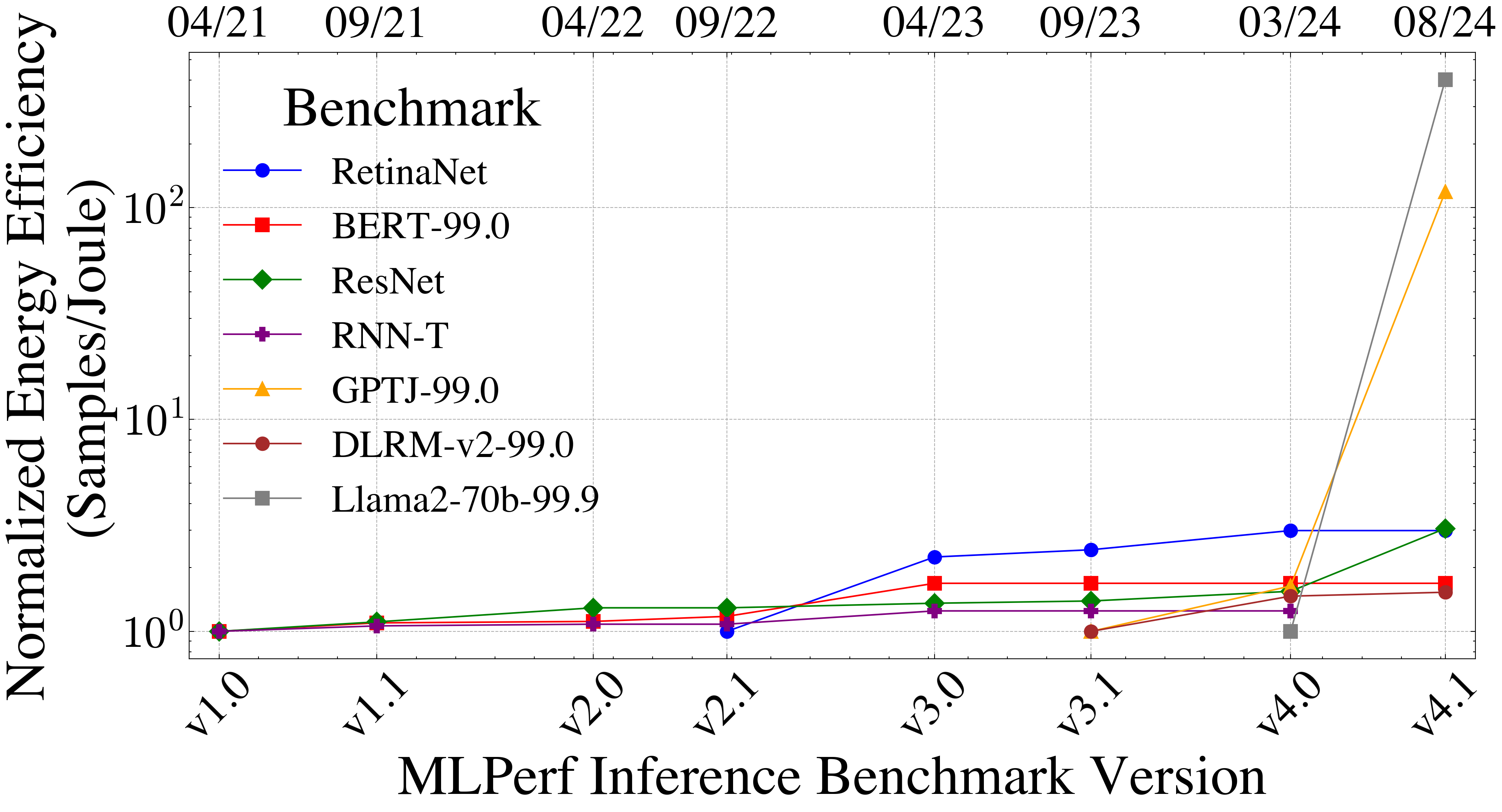}
        \caption{Datacenter}
        \label{fig:dc_trends}
    \end{subfigure}
    \hfill
    \begin{subfigure}[b]{0.32\textwidth}
        \centering
        \includegraphics[width=\textwidth]{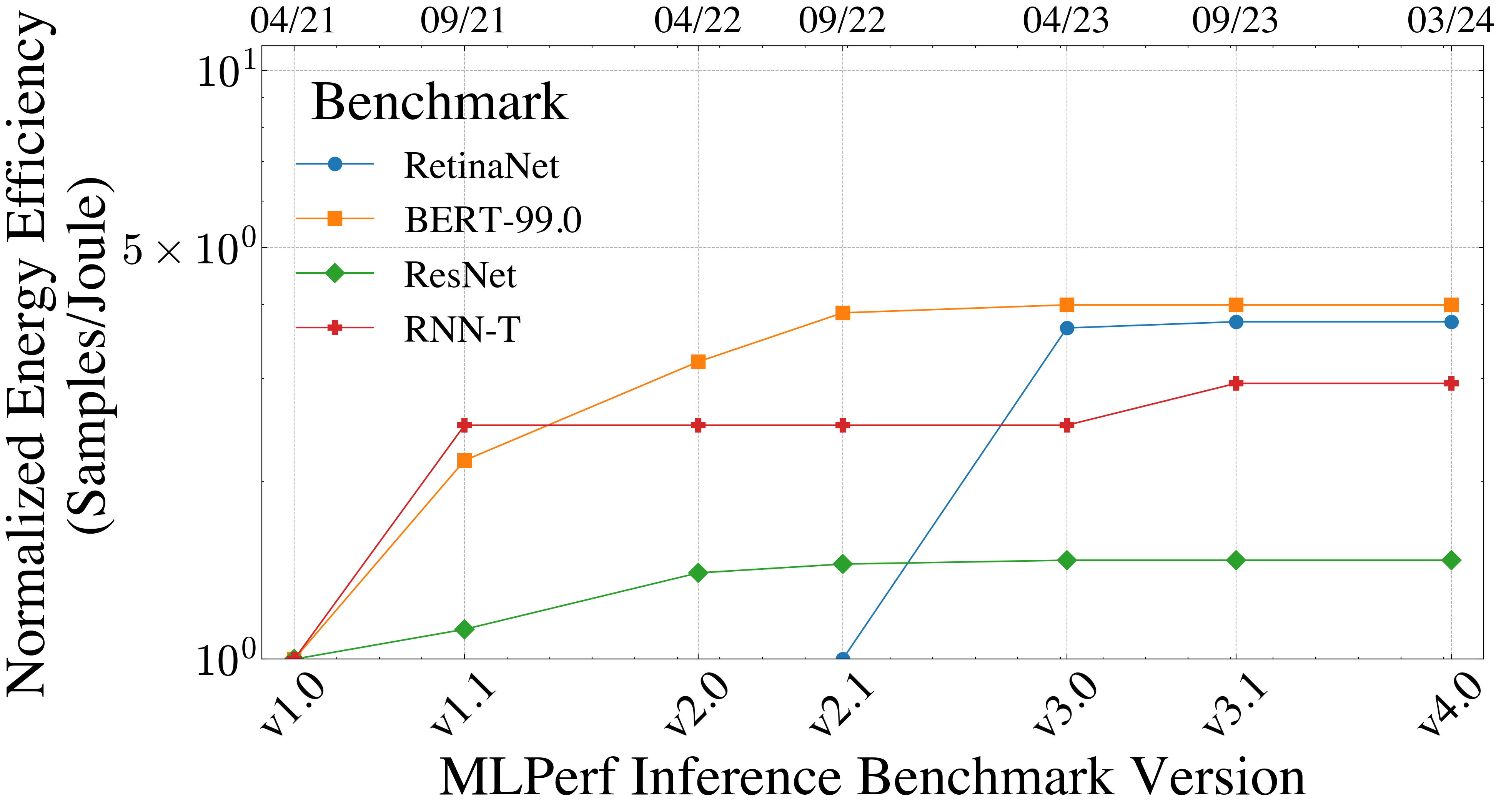}
        \caption{Edge}
        \label{fig:edge_trends}
    \end{subfigure}
    \hfill
    \begin{subfigure}[b]{0.32\textwidth}
        \centering
        \includegraphics[width=\textwidth]{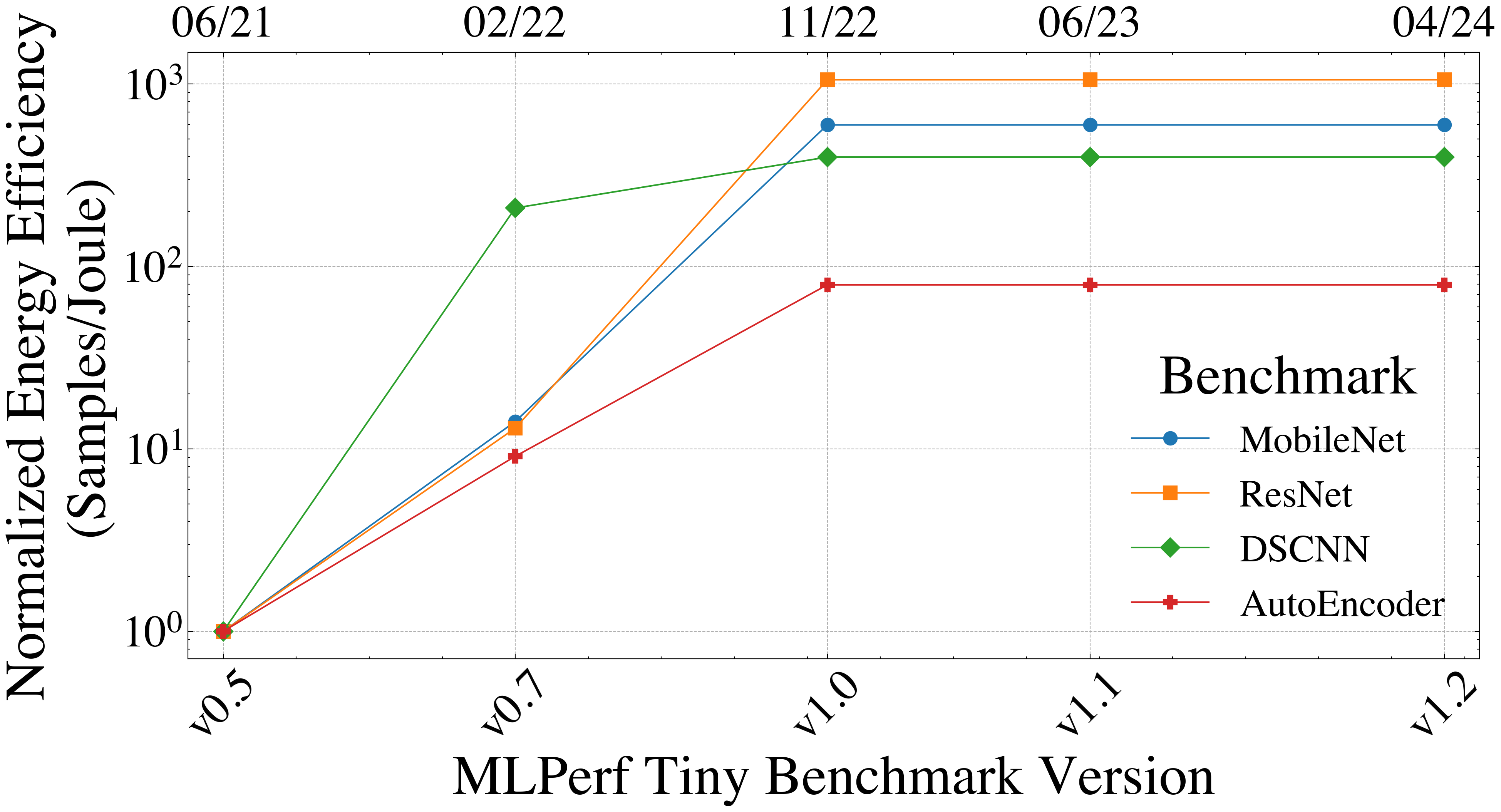}
        \caption{Tiny}
        \label{fig:tiny_trends}
    \end{subfigure}
    \caption{Comparison of energy efficiency trends for datacenter, edge, and tiny inference.}
    \label{fig:comparison}
    \vspace{-5mm}
\end{figure*}



We analyze this diverse range of data to gain an understanding of comparative trends across different workloads and system scales, shedding light on the areas of focus within the industry. MLPerf benchmarks span a diverse range of ML workloads, from resource-constrained keyword spotting to complex computer vision, sophisticated recommendation systems, and state-of-the-art large language models (LLMs)~\cite{mlperf, mlcommons}. With each version, we make changes to the MLPerf benchmarks to adapt to the changing ML landscape, such as updating datasets, adding new benchmarks, updating the network architectures, and refining our methodology and hyperparameters \cite{mlperfvision}. A notable example is the addition of LLMs in version 3.1 to address the growing interest in this area.

The MLPerf Power submissions offer a large dataset that spans several years and four workload versions. This extensive collection, comprising 1,841 submission results, provides useful insight into industry-wide trends in energy efficiency improvements. The data set includes 590 data center results~\cite{datacenterdata}, 792 edge results~\cite{edgedata}, 447 tiny MLPerf results~\cite{tinydata} for inference, and 12 training submission results~\cite{trainingdata}. Unless otherwise specified, all results are verified MLPerf submissions~\cite{mlperfmessaging}. Configurations (performance-optimized, energy efficiency-optimized, or default) and logs of individual MLPerf submissions are publicly available on Github.

\subsection{Overall Energy Efficiency Improvements in Inference}
\label{sec:efficiency_gains}

Following our earlier examination of performance trends shown in Figure~\ref{fig:mlperf_performance}, we now shift our focus to advancements in AI energy efficiency. Figure~\ref{fig:comparison} highlights significant improvements across the categories of data center, edge, and tiny devices. These advancements are expressed using normalized samples per joule, a measure that encapsulates enhancements in processing power alongside an industry-wide emphasis on optimizing AI technology for energy efficiency.

In the data center category, \texttt{RetinaNet}~\cite{retinanet} shows the strongest increase in samples per joule amongst older benchmarks. However, it is clear to see that the new generative AI benchmarks exhibit massive energy efficiency improvements in recent versions. \texttt{GPT}~\cite{gpt3} and \texttt{Llama2}~\cite{llama2} both reach over 100$\times$ improvements in energy efficiency, which can be attributed to the significant attention given to LLM computing efficiency as a result of its significant impact on scale. Overall, data center models show large gains in energy efficiency, with notable improvements in the later iterations.

In the edge category, improvements in energy efficiency are moderate. \texttt{BERT} at 99\% accuracy, or \texttt{BERT-99.0}, and \texttt{RNN-T} are at the forefront with substantial early improvements, reaching up to four times their initial efficiency. In contrast, \texttt{ResNet} shows a more steady enhancement, leveling at approximately 1.5 normalized samples per joule. Progress in edge devices underscores the emphasis on refining AI models for real-time, on-device execution, crucial for applications that demand low latency and high energy efficiency.

In the tiny category, we observe significant improvements in energy efficiency across every model. \texttt{ResNet}~\cite{resnet1} achieves a normalized efficiency gain over 1000$\times$, while the other 3 workloads all reach between 79$\times$ and 596$\times$. These results underscore the rapid advances in AI efficiency for small-scale, resource-constrained environments. In particular, tiny-scale computing has experienced massive improvements in energy efficiency over its early versions, with gains between 2 and 3 orders of magnitude across every workload. 

The trends across all three categories highlight the ongoing efforts and successes in enhancing AI models' energy efficiency, ensuring that high performance can be achieved sustainably across different environments and use cases. However, for edge and tiny systems, energy efficiency advancements have plateaued in the last 3 MLPerf versions, indicating a need for more innovative approaches to continue improving efficiency. Improving their energy efficiency remains crucial for the scalability and sustainability of future IoT systems.

\subsection{Scalability Analysis of Training}


As ML models grow in complexity, the time to solution (train) becomes a critical factor in the design of the ML system~\cite{goldenage, dawnbench}. Researchers and industry practitioners have developed various parallelization strategies to reduce training time by leveraging additional computational resources. However, our MLPerf Power benchmarking reveals that this pursuit of faster training times introduces a new dimension of complexity: energy consumption at scale. To this end, we present a novel analysis of the intricate relationship between system scale, training time, and energy consumption, drawing from our recently released MLPerf Training power benchmarks. 

\begin{figure}[t!]
\centering
\includegraphics[width=\columnwidth]{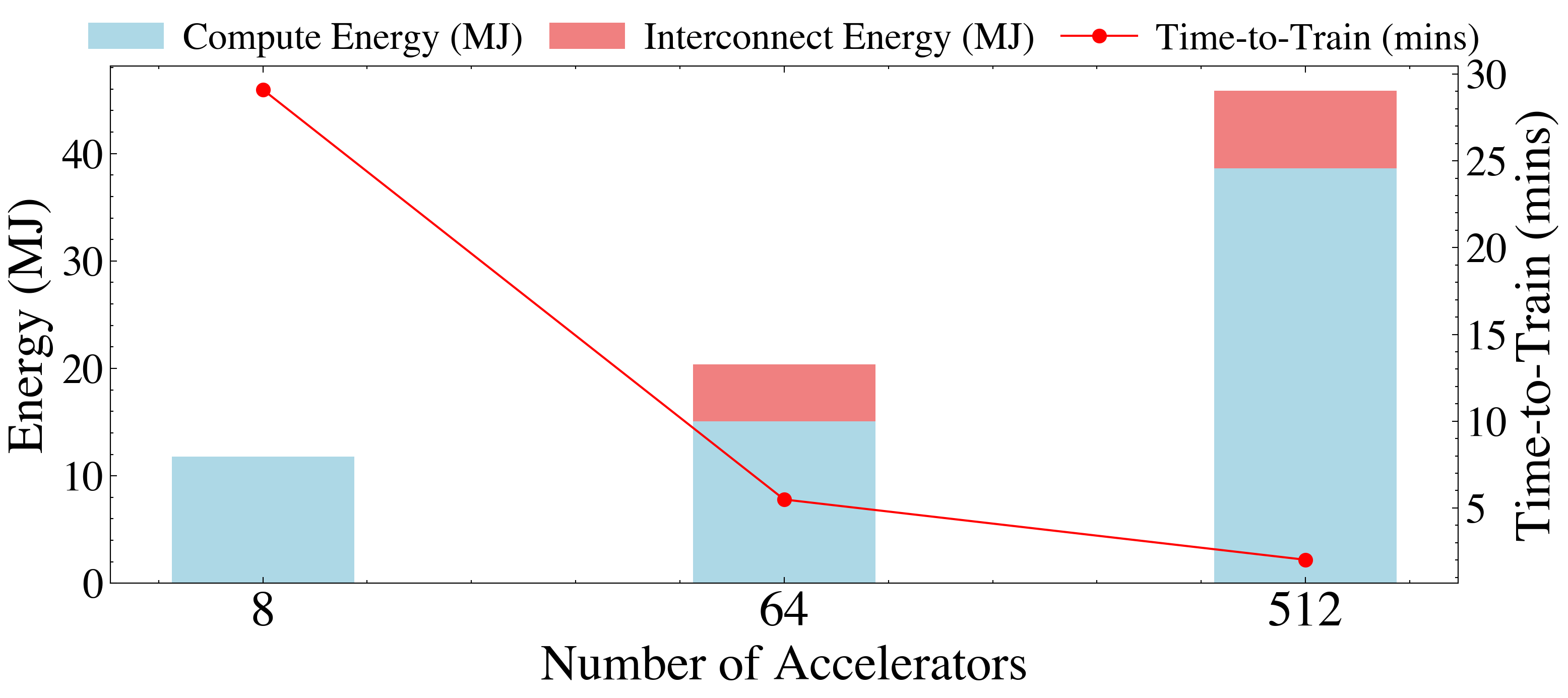}
\caption{Energy consumption and time-to-train for Llama2-70b LoRA fine-tuning across different numbers of accelerators.}
\label{fig:training_energy_breakdown}
\end{figure}

Figure \ref{fig:training_energy_breakdown} shows the time-to-train and energy consumption metrics from our v4.0 training power submissions for the \texttt{Llama2-70b}~\cite{llama2} model on three distinct system scales. This data provides insights into the non-linear nature of energy scaling in large-scale ML training systems. With perfect scaling, increasing the number of accelerators should lead to a proportional reduction in training time and the energy to train should remain flat. But our results reveal a complex reality. 

As the system scale increases, we observe diminishing returns in training time reduction. The improvement in training time becomes less pronounced as we add more accelerators, due to increased inter-accelerator communication and reduced FLOPs utilization in each accelerator.  This results in higher accelerator-hours to train despite absolute time-to-train being lower. Simultaneously, we see an increase in energy to train. This is due to the increase in total accelerator-hours to train and the added energy cost of interconnect components, such as networking switches, required for inter-node communication.


Our analysis suggests several key implications for future ML system design. First, architects must carefully consider the energy costs associated with scaling up training systems, which can lead to the development of more energy-efficient accelerators and interconnects. This energy-aware scaling approach is crucial for sustainable AI development. 

Second, understanding the parallelization characteristics of different ML workloads can help to determine the optimal system scale that balances performance gains with energy efficiency. This workload-specific optimization will be key to maximizing efficiency across various ML tasks. Lastly, future ML systems should optimize not just compute energy, but also other types of energy consumption, such as interconnect energy, which becomes increasingly significant at larger scales. 

\subsection{Workload-Specific Insights}

Unlike raw power measurements, which are impacted by how submitters scale their hardware in relation to model size, energy consumption is agnostic to constant power or latency requirements. For this reason, we use the energy per benchmark inference to evaluate its computational demands and efficiency. We provide insights into the relationships between model size, workload characteristics, and energy consumption of each benchmark submission by evaluating two divisions of inference benchmarks, datacenter and tiny. Note that while there is no perfectly comparable metric for model size, we use total matrix accumulation (MAC) operations as an estimate. For an in-depth description of the workloads and deeper insight into the computation of each benchmark, refer to the MLPerf-Inference \cite{mlperfinference} and MLPerf-Tiny \cite{mlperftiny} papers.



\begin{figure}[t!]
\centering
\includegraphics[width=\columnwidth]{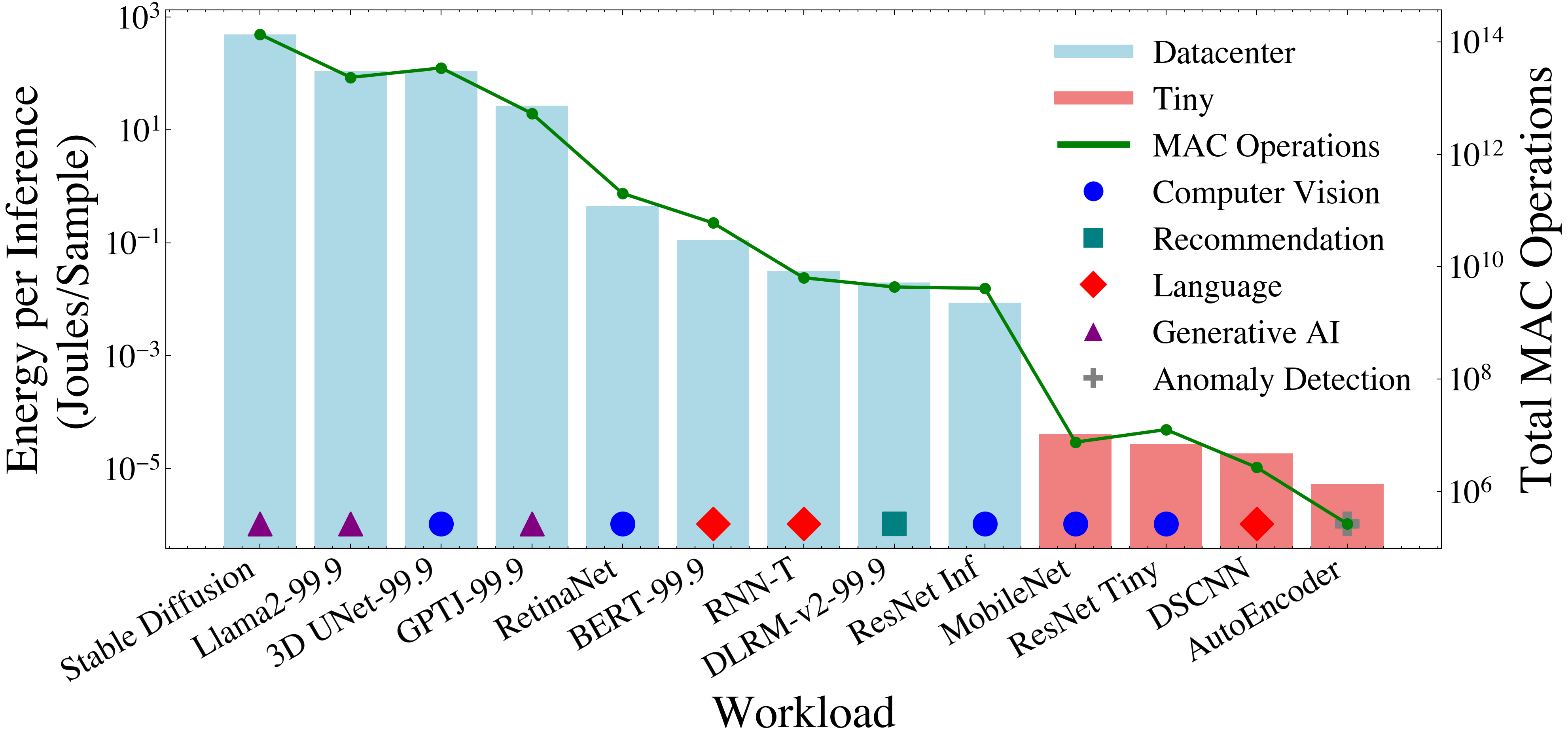}
\caption{Energy consumption and total MAC operations per inference for each MLPerf Inference and MLPerf Tiny benchmarks. Each benchmark is categorized by its workload type.}
\label{fig:efficiency_workloads}
\end{figure}

\textbf{Datacenter} The blue bars in Figure \ref{fig:efficiency_workloads} show the energy per inference sample and workload computation size for each MLPerf Inference benchmark on a large-scale datacenter system. \texttt{ResNet}~\cite{resnet1, resnet2}, a classic image classification model widely used as a baseline for computer vision (CV) tasks, consumes the least energy per inference at 8.7 mJ/Sample. As computer vision models get larger and more capable, the energy for a single inference also grows. \texttt{RetinaNet}~\cite{retinanet} is an object detection model that uses multiple \texttt{ResNet} inferences for multiple classifications within a single image and thus a multiple order-of-magnitude growth in inference energy. The computer vision benchmark, \texttt{3D U-Net}~\cite{3dunet} is a 3-dimensional medical image segmentation model that uses even more energy per inference due to the extra visual dimension and required precision.

Compared to CV models, recommendation models such as \texttt{DLRM-v2}~\cite{dlrm} exhibit very different data movement patterns~\cite{hildebrand2023efficient}. Recommendation models are known to have memory bandwidth bottlenecks and therefore exert a greater percentage of total energy on data movement than compute bound models. Although \texttt{DLRM-v2} only has 1.06$\times$ more MACs than \texttt{ResNet}, the 2.3$\times$ increase in J/Sample can be attributed to the data movement energy not captured in pure computation measurements. 


We also evaluate two NLP benchmarks. \texttt{RNN-T}~\cite{rnnt} is a speech recognition model designed to transcribe spoken language into text, while \texttt{BERT}~\cite{bert} is a multipurpose encoder-only transformer language model capable of various tasks such as question answering. Despite their model architecture similarities, these workloads likely have much lower inference energy requirements than an LLM like \texttt{GPT-J} due to 1 order of magnitude fewer parameters and and 2 orders of magnitude fewer MAC computations.

The three generative AI models in MLPerf Inference are significantly more energy-hungry than every benchmark other than \texttt{3D U-Net}. Generative transformer-based model architectures are much more complex than other models in both raw MAC computation and data movement~\cite{keles2023computational}. \texttt{Stable Diffusion}~\cite{stable-diffusion} computes 673$\times$ more MACs and consumes 1082$\times$ more energy for a single image generation inference compared to \texttt{RetinaNet}'s image object detection network. LLM inference is the most widely used generative AI model in industry and is benchmarked by the 70B parameter \texttt{Llama2}~\cite{llama2} and the 6B parameter \texttt{GPT-J}~\cite{gpt3}. Interestingly, despite the 11.7$\times$ increase in parameters, \texttt{Llama2} computes 4.4$\times$ more MACs and uses 4.2$\times$ more energy per inference than \texttt{GPT-J}. This discrepancy could be due to the non-linear scaling of energy efficiency in language models by parameters, better energy efficiency of \texttt{Llama2}, or differences in the benchmark datasets.

It is important to note that autoregressive language models exhibit variability in their inference lengths. In MLPerf Inference v4.0, each of the two LLM benchmarks uses a different metric. \texttt{GPT-J} is reported in samples/second, while \texttt{Llama2} is given in tokens/second. From version 4.1 onward, all LLM benchmarks will be reported in tokens/second, as is common in industry. However, for better alignment with how humans interact with LLMs and comparability with other MLPerf Inference benchmarks, we use samples/second to evaluate LLM performance for energy efficiency calculations in Fig. \ref{fig:efficiency_workloads}. To do this, we convert tokens into samples using an estimated median sequence length of 69 tokens/sample for \texttt{GPT-J} and 292 tokens/sample for \texttt{Llama2}, with an error margin due to sequence length variation, user pairs, and other factors.

As newer ML models grow in their computational complexity, their energy consumption scales by multiple orders of magnitude. While our largest benchmarked LLM \texttt{Llama2} consumes 111.4 J/Sample at 70B parameters, current models are growing into the multi-trillion parameter scales \cite{hudson2023trillion}. Due to propriety and compute resource limitations, we cannot explicitly benchmark these large language models in MLPerf. However, it is clear that as energy consumption reaches multi-KJs for a single question-answer inference, it is more important than ever to consider energy when designing and optimizing models for more complex tasks.

\textbf{Tiny} Similar trends are seen in MLPerf-Tiny. \texttt{AutoEncoder}~\cite{autoencoder1, autoencoder2, autoencoder3}, an anomoly detection model, consumes the least energy at 20 mJ per inference on the reference hardware and computes the fewest MACs at 265k (without batch norm operations). \texttt{MobileNet}~\cite{mobilenet}, a visual wake-word detection model, and \texttt{DSCNN}~\cite{dscnn}, a keyword-spotting model, follow similar trends with model complexity leading to higher energy per inference.

CV with \texttt{ResNet}~\cite{resnet1} is an interesting comparison between datacenter and tiny systems. The datacenter \texttt{ResNet} benchmark uses the 224x224 ImageNet dataset at 99\% inference accuracy, while the tiny \texttt{ResNet} uses the 32x32 CIFAR10 dataset at 85\% inference accuracy. With the 49$\times$ larger input dataset, 14\% increase in the accuracy target and 326$\times$ increase in the total MACs computed, the datacenter system consumes 321.7$\times$ more energy per inference than the tiny system. While ML ASICs in datacenter scale systems and embedded neural processors in tiny scale systems encounter different energy bottlenecks, there is strong continuity between workload complexity and inference energy across hardware scales. 










\subsection{Energy Efficiency Costs of High Accuracy Inference}

Achieving high accuracy in an ML model generally requires more computational resources due to the increased complexity in calculations, data processing, and additional training iterations. Each MLPerf benchmark includes a minimum accuracy target that must be achieved to qualify for a successful submission. Several MLPerf workloads have separate benchmarks set at 99\% and 99.9\% accuracy. For example, in the low accuracy \texttt{BERT-99.0} benchmark, participants must achieve 99\% of the original model's single precision FP32 accuracy.

\begin{figure}[]
    \centering
    \includegraphics[width=\columnwidth]{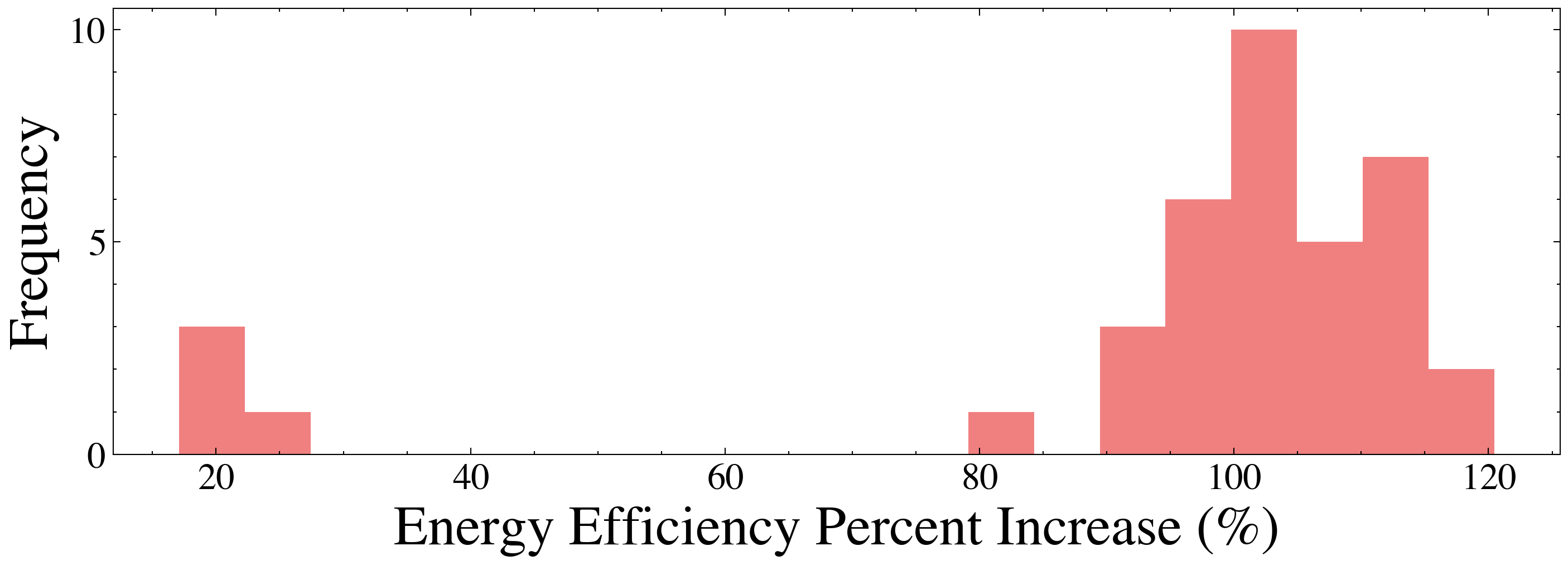}
    \caption{Distribution of BERT energy efficiency drop when increasing accuracy from 99\% to 99.9\% for datacenter offline submissions in MLPerf v3.1, v4.0, and v4.1.}
    \label{fig:accuracy_efficiency}
\end{figure}

Figure \ref{fig:accuracy_efficiency} shows the energy efficiency cost distribution on the BERT benchmark as accuracy increases from 99\% to 99.9\% in offline datacenter submissions for versions 3.1 and 4.0. Energy efficiency cost is measured as the percentage change in Samples/Joule when transitioning from \texttt{BERT-99.0} to \texttt{BERT-99.9}. We observe that datacenter systems running inference at 99.9\% accuracy are, on average, half as efficient as those running inference at 99\% accuracy. However, the three samples on the left showcase the capabilities of recent submissions that utilize more advanced quantization techniques to maintain strong energy efficiency at higher accuracy. This trend indicates that running inference at higher accuracy is becoming cheaper and more sustainable.



\subsection{Low Precision Efficiency Improvement Analysis}

\begin{figure}[t!]
    \centering
    \includegraphics[width=\columnwidth]{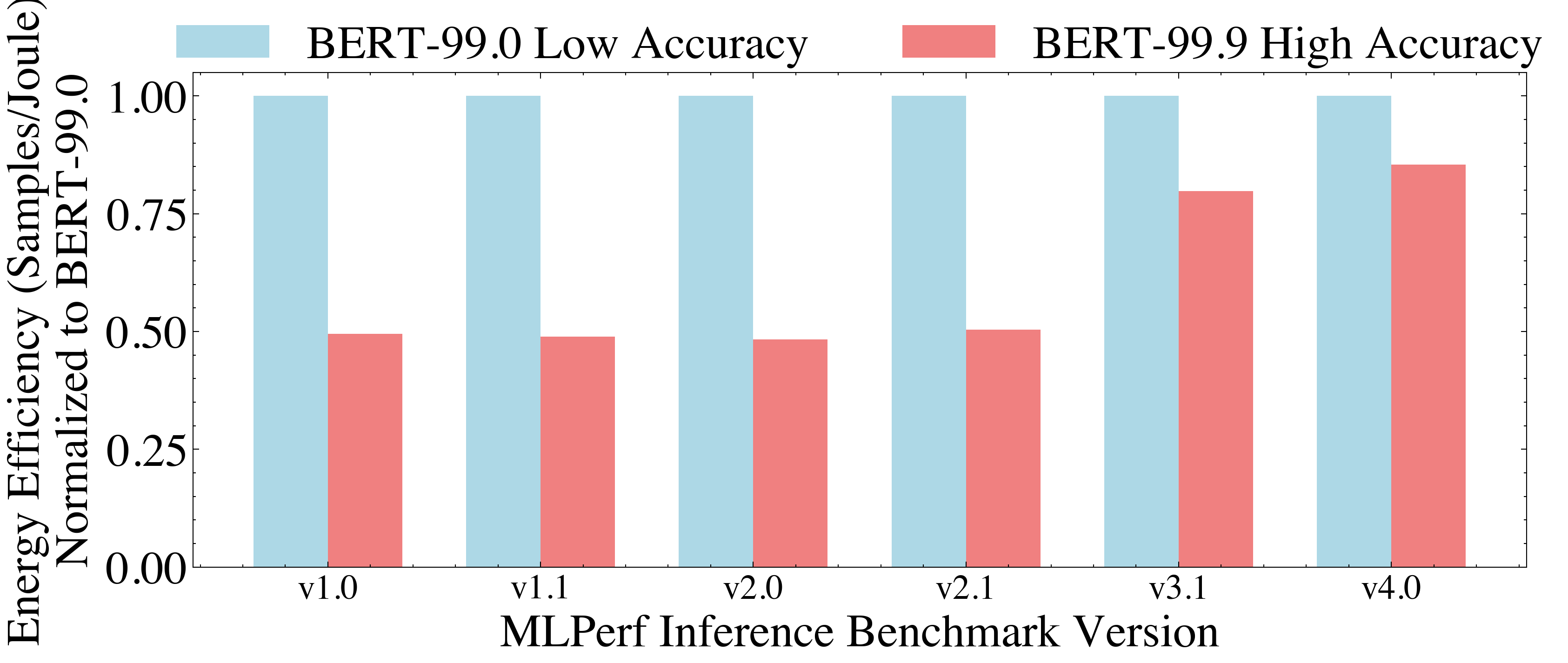}
    \caption{Energy efficiency for high and low accuracy targets for \texttt{BERT} MLPerf-Inference. INT8 quantization is used for \texttt{BERT-99.0} submissions, while \texttt{BERT-99.9} submissions initially used FP16 and later adopted FP8 starting from v3.1.}
    \label{fig:quant_improvement}
\end{figure}

Using quantization improves performance~\cite{quantperf1, quantperf2} and reduces power usage ~\cite{quantenergy1, quantenergy2} by decreasing the amount of computation, either directly or by enhancing parallelism. However, due to the stringent accuracy standards set by the MLPerf benchmarks, participants must find a careful balance between aggressive quantization to boost performance and energy efficiency while still meeting the accuracy criteria. Figure \ref{fig:quant_improvement} examines the normalized energy efficiency of systems using workloads \texttt{BERT-99.0} and \texttt{BERT-99.9} on the same hardware platform within the same MLPerf-Inference version.

\texttt{BERT-99.0} submissions consistently use INT8 quantization, as it is sufficient to meet the 99\% accuracy target. However, INT8 is inadequate for the higher accuracy target of \texttt{BERT-99.9}, which requires the use of 16-bit floating point (FP16) until version 3.1. Starting from v3.1, the new hardware that supports 8-bit floating point (FP8) formats allowed submitters to achieve higher performance and energy efficiency while still meeting the 99.9\% accuracy target. Consequently, the energy efficiency of \texttt{BERT-99.9} submissions improved significantly from approximately 50\% of \texttt{BERT-99.0} in older submissions to around 85\% in newer ones. 

Thus, the adoption of lower precision floating point formats, such as FP8, leads to reduced energy consumption per operation and energy usage in memory data transfers. Our analysis strongly suggests that innovation in quantization techniques will continue being a driving force in energy efficient systems.

\subsection{Software versus Hardware-driven Efficiency Improvements}

As discussed in Section~\ref{sec:efficiency_gains}, while significant improvements in efficiency have been observed in the past, recent versions of MLPerf show signs of plateauing (Figure~\ref{fig:comparison}). We analyze these advancements by isolating the sources of efficiency gains, measuring their extent, and evaluating their pros and cons. Our goal is to provide hardware and software engineers with a better understanding of the origins of energy enhancements to enable  more informed decision making. 



\begin{figure}[t!]
    \centering
     \includegraphics[width=\columnwidth]{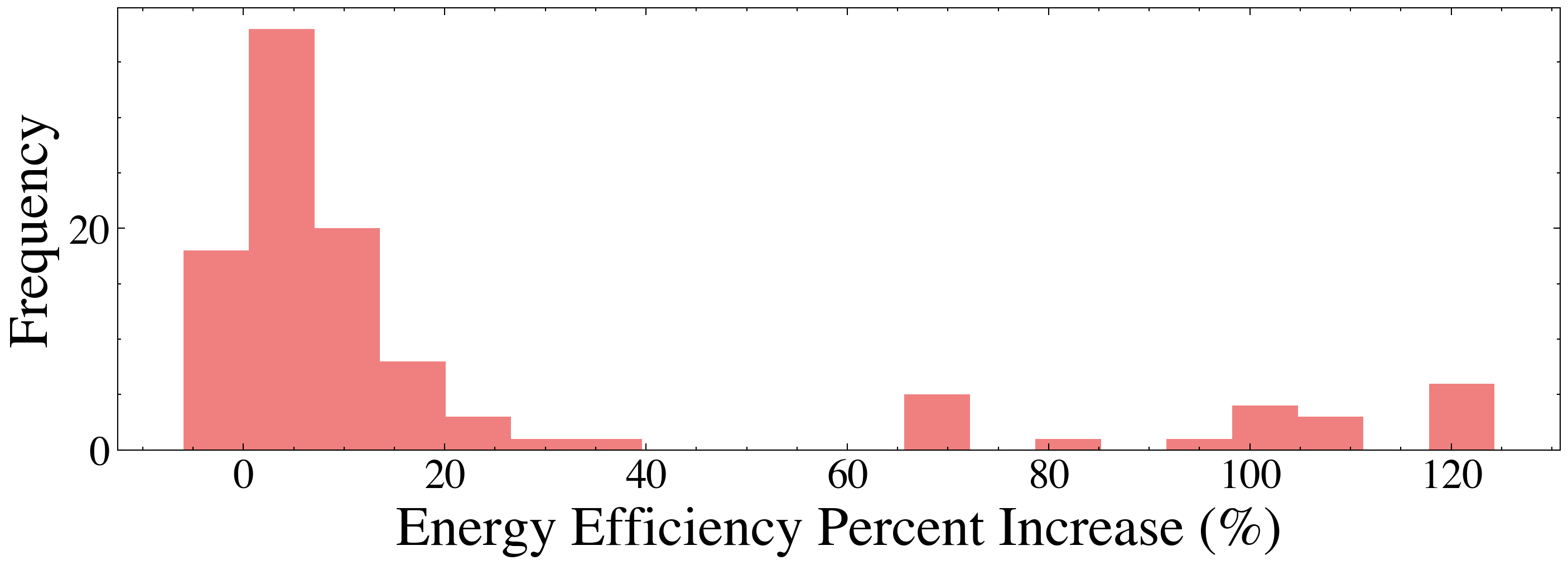}
    \caption{Distribution of energy efficiency change when optimizing only the software between identical hardware system submissions in consecutive MLPerf datacenter offline versions.}
    \label{fig:software_histogram}
\end{figure}

Figure \ref{fig:software_histogram} illustrates submissions with identical hardware configurations (host CPU, number of accelerators, type of accelerator) in two successive versions of any workload. The histogram shows that 89\% of the data is positive, suggesting that most submissions on similar hardware have some degree of improvement in software efficiency. We propose that the small percentage of data showing a minor decline in efficiency might be due to configuration changes not considered in our hardware equivalency or vendor modifications that do not focus on energy efficiency. Furthermore, it is notable that 18\% of the data achieves improvements in energy efficiency that exceed 50\%, providing strong evidence that software optimizations have significantly improved energy efficient systems.

\begin{figure}[t!]
    \centering
    \begin{subfigure}[b]{\columnwidth}
        \centering
        \includegraphics[width=\textwidth]{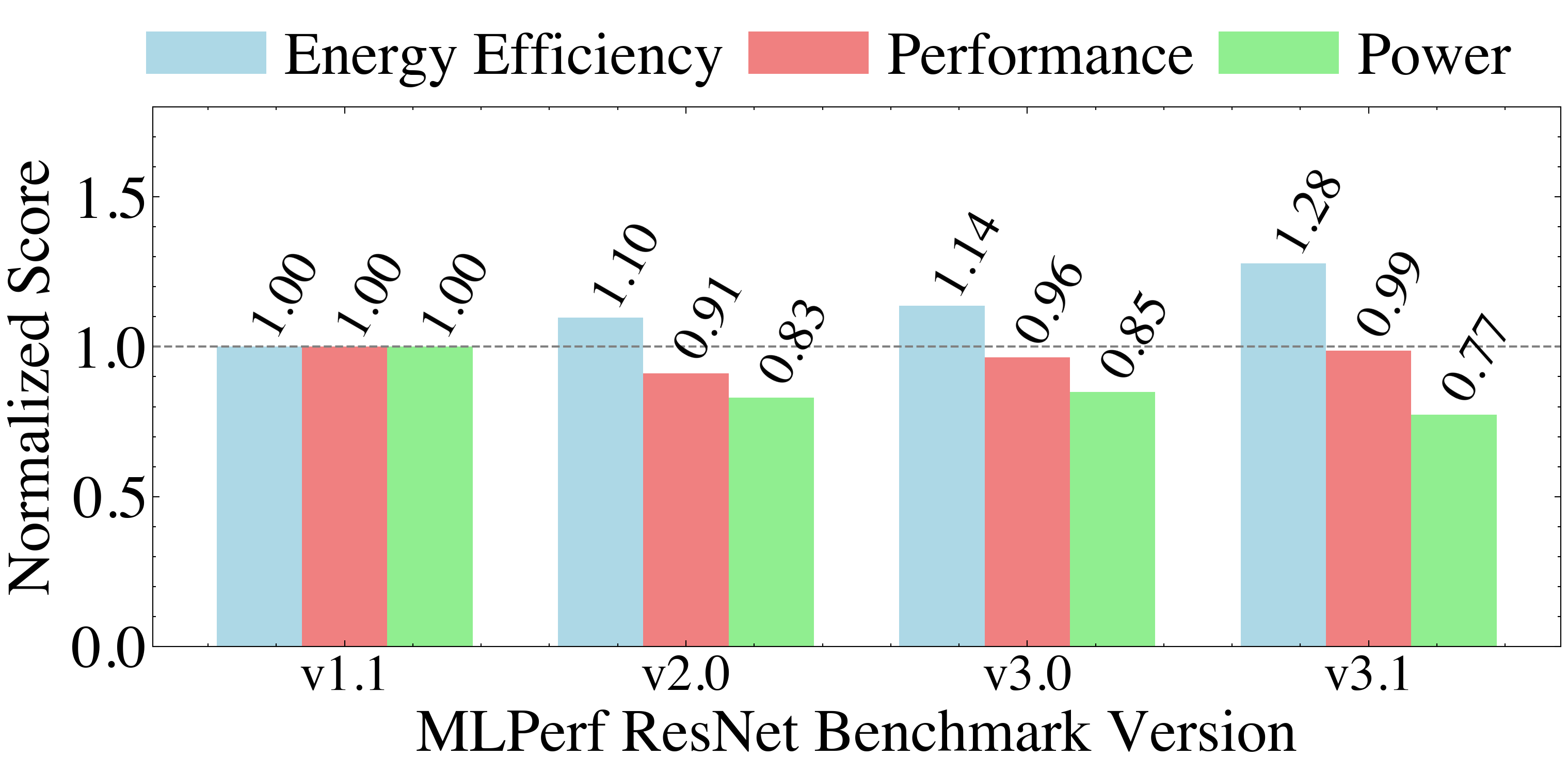}
        \caption{Software Optimizations}
        \label{fig:software_improvement}
    \end{subfigure}
    \begin{subfigure}[b]{\columnwidth}
        \centering
        \includegraphics[width=\textwidth]{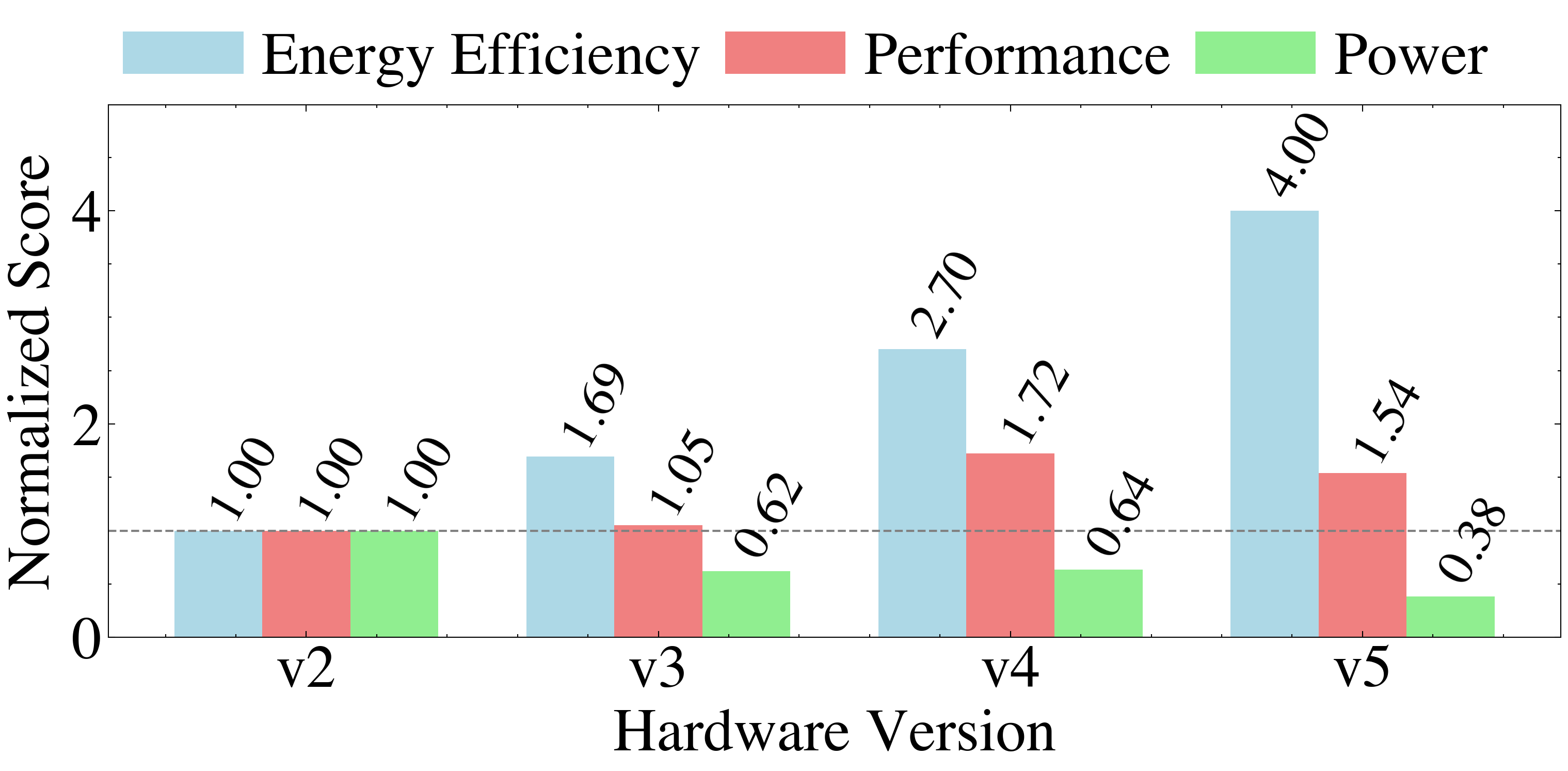}
        \caption{Hardware Optimizations}
        \label{fig:hardware_improvement}
    \end{subfigure}
    \caption{Normalized energy efficiency, performance, and power improvements from progressive versions of software-isolated and hardware-isolated optimizations.}
    \label{fig:combined_improvements}
\end{figure}

We analyze a case study on software enhancements in Figure \ref{fig:software_improvement}, identifying an identical edge system in four consecutive versions of \texttt{ResNet} inference workloads. With negligible performance loss, power consumption has decreased at a fast pace, resulting in 1.28$\times$ energy efficiency improvement. Submission logs show that improvements in software development kits (SDKs) and total cost of ownership (TCO) contribute to the gains in energy efficiency. Sacrificing a small amount of performance for these software-level optimization additions results in energy efficiency enhancements.

In addition to software improvements, we see improvements in hardware to reduce training time and energy. In Figure \ref{fig:hardware_improvement}, we evaluate a case study on hardware where we use a constant software stack, including quantization, and evaluate subsequent versions of the ASIC on the \texttt{BERT} benchmark.\footnote{Figure \ref{fig:hardware_improvement} results are unverified by MLPerf Power. This data is used here only to show general trends in energy efficiency from hardware optimizations.} Performance increases to 1.72$\times$ and power consumption decreases to 0.38$\times$, giving a 4$\times$ increase in energy efficiency in the latest hardware version of the system. The computing units in newer hardware versions have features that yield better energy efficiency: support for lower numeric precision, more efficient designs for the same precision, and better handling of memory-compute coordination~\cite{tpuv1, tpuv4}. We also see improvements in the fast interconnects between ML accelerators in terms of higher bandwidth, lower latency, and support for more diverse communication patterns. Furthermore, power management techniques such as dynamic voltage and frequency scaling allow processors to adjust their power consumption based on workload demands. Our goal in analyzing these advancements is to provide architects and ML engineers with a better understanding of the origins of energy enhancements to enable better and more informed decision making.


\section{Recommendations for Future Directions}

We have identified four main areas that require the attention of the ML systems research community and industry.

\textbf{Mitigating Efficiency Plateau.} Our analysis quantitatively shows that significant improvements in energy efficiency in new generative AI benchmarks in the datacenter category due to spiking commercial interest in its performance and energy efficiency. However, recent MLPerf versions are indicating a plateau in advancements in the edge and tiny categories. To overcome this, we recommend developing granular, workload-specific energy efficiency metrics that account for unique model characteristics, training regimes, and energy cost per accuracy improvement. We also suggest analyzing the energy efficiency of individual training and inference phases. 

\textbf{Promoting AI Sustainability.} We recommend integrating environmental impact considerations into AI development practices and and ethical frameworks. Beyond including energy efficiency as a key metric, we recommend developing and standardizing tools to estimate the carbon footprint of AI model training and inference, based on existing work in this area. Existing~\cite{codecarbon,carbonresponder,carbondependencies, carbonexplorer, clover, sustainablecloud} and new tools should be integrated into AI frameworks to provide real-time feedback on power consumption during the development process. Furthermore, we suggest exploring the creation of incentive structures, such as energy efficiency certifications or awards, to encourage energy-efficient ML system design. 


\textbf{Embracing AI Systems Policy} Our work directly supports emerging regulations like the European Union AI Act’s Article 53~\cite{eu_ai_act53}, which requires AI model providers to document energy usage. The MLPerf Power benchmarking methodology can help meet these regulatory demands by providing standardized and reproducible energy efficiency measurements. As sustainability becomes a key focus in AI, this benchmark will be crucial for promoting transparency and compliance. Future research should explore integrating these benchmarks into regulatory frameworks~\cite{sustainable_policy1, sustainable_policy2, emissions} and expanding the policy to cover new AI workloads and hardware systems.

\textbf{Nurturing Industry-Wide Collaboration.} The diverse nature of AI workloads and rapid pace of technological advancement require continued coordinated efforts to drive improvements in energy efficiency. Building on the existing work of MLCommons~\cite{mlcommons} and its MLPerf Working Groups~\cite{powerWG, trainingdata, inferenceWG, tinydata}, we recommend to growing these collaborations to address emerging challenges in AI energy efficiency. In addition to expanding the scope of current benchmarks to include more diverse and emerging AI workloads, such as computer vision~\cite{co2cv}, multi-model~\cite{xiao2018deep, shen2019meal} and multimodal~\cite{ramachandram2017deep, ngiam2011multimodal} workloads, we suggest developing guidelines for reporting the environmental impact of AI model development and deployment that complement the energy efficiency metrics.

\textbf{Large-Scale AI Training Optimization.} Our analysis of \texttt{Llama2-70b} fine-tuning, reveals a complex relationship between system scale, training time, and energy consumption. We recommend developing granular, workload-specific energy efficiency metrics for large-scale AI training that account for the unique characteristics of different AI models and training regimes, going beyond overall system efficiency to examine component-level energy consumption patterns. We also recommend creating a framework for analyzing the energy efficiency of individual training phases, such as data loading, forward passes, backward passes, and parameter updates. This granular approach could help identify specific bottlenecks in large-scale training systems and guide targeted optimizations. Additionally, we suggest incorporating scaling efficiency metrics that quantify the energy cost of distributed training.

\textbf{Accuracy-Efficiency Trade-offs.} Our analysis of the \texttt{BERT} and \texttt{DLRM-v2} benchmarks demonstrates the significant impact of quantization on energy efficiency and reveals substantial energy costs associated with high-accuracy inference. To address the accuracy-efficiency trade-off more explicitly, we recommend extending the MLPerf Power framework to include fine-grained metrics that quantify energy cost per accuracy improvement and analyze the submission results.

Additionally, incorporating methodologies for tiered accuracy offerings would enable dynamic adjustment of model accuracy based on application requirements and energy constraints~\cite{halpern2019one,romero2021infaas}. These enhancements would provide a more comprehensive framework for evaluating and optimizing the energy efficiency of AI models across various accuracy levels and deployment scenarios, ultimately guiding the development of more energy-efficient AI systems.

\textbf{Multimodal Deployments.} As part of MLCommons' ongoing effort to add a benchmark for the audiovisual (AV) domain~\cite{zhu2021deep}, we are incorporating multiple multimodal (camera + LiDAR, for example) neural networks covering key elements of the AV stack which will execute concurrently.  As power consumption is primary concern for such use cases, we are currently devising an appropriate methodology for power measurement.

\textbf{Software-Hardware Co-design Paradigm.} As we show, software optimizations alone can lead to significant energy efficiency improvements. We recommend establishing formal collaborative frameworks for software-hardware co-design in AI system development to facilitate early and ongoing interaction between hardware designers and software engineers, with shared energy efficiency targets guiding development processes.
\section{Conclusion}

MLPerf Power introduces a standardized methodology for benchmarking energy efficiency in machine learning systems across an unprecedented scale, from microwatts to megawatts. This work addresses the critical need for understanding and optimizing power consumption in ML as the field advances rapidly. Our key contributions include establishing an industry-wide standard for ML system power measurement, presenting the first large-scale study of ML system energy efficiency, analyzing longitudinal energy efficiency trends across multiple generations of MLPerf, and providing actionable insights on energy optimization techniques. Our findings emphasize the importance of energy efficiency as a primary metric in ML system design. As the field progresses, the insights gained from MLPerf Power will serve as a stepping stone towards the future of sustainable machine learning systems.


\section*{Acknowledgements}

MLPerf Power is the collective effort of numerous individuals from various organizations. In this section, we would like to acknowledge all those who contributed to producing the initial set of results or supported the overall benchmark development. This work was supported in part by funding from SRC and NSF.\\

\noindent \textbf{Broadcom} - Ravi Soundarajan. \\
\noindent \textbf{Databricks} - Hanlin Tang. \\
\noindent \textbf{Fujitsu} - Takahiro Notsu. \\
\noindent \textbf{Google} - Tom Jablin, David Patterson. \\
\noindent \textbf{Infineon} - Peter Torelli. \\
\noindent \textbf{Intel} - Remya Jayaraman. \\
\noindent \textbf{KRAI} - Leo Gordon. \\
\noindent \textbf{Meta} - Carole-Jean Wu, Whitney Zhou. \\
\noindent \textbf{MLCommons} - Pablo Gonzales Mesa. \\
\noindent \textbf{NVIDIA} - Ashwin Nanjappa, Ashutosh Dhar, Dilip Sequeira. \\
\noindent \textbf{Xored} - Albert Safin, Yaroslav Kurlaev, Julia Mozhar, Daniil Efremov, Andrey Atangulov, Sergei Lunin.

\bibliographystyle{IEEEtranS}
\bibliography{refs}

\end{document}